\documentclass[12pt,a4paper]{article}
\usepackage[top=1.52in, bottom=1.52in, left=2in, right=2in]{geometry}
\usepackage{amssymb,enumitem,booktabs,subfig,xcolor,rotating,todonotes,natbib,bm,dsfont,paralist,setspace,tikz,paralist}
\usepackage[pdftex,colorlinks=true]{hyperref}
\definecolor{darkblue}{rgb}{0,0,.6}
\hypersetup{citecolor=darkblue,linkcolor=darkblue,urlcolor=darkblue}
\usepackage{amsmath}
\usepackage{amsfonts}
\usepackage{epsfig}
\usepackage{graphics}
\usepackage{graphicx}

\providecommand{\U}[1]{\protect\rule{.1in}{.1in}}

\DeclareMathOperator*{\argmin}{\arg\!\min}
\graphicspath{{plots/}}
\newsavebox\CBox
\def\textBF#1{\sbox\CBox{#1}\resizebox{\wd\CBox}{\ht\CBox}{\textbf{#1}}}

\spacing{1.25}

\begin{document}

\begin{center}
\large Functional time series forecasting with dynamic updating: \hbox{An application to intraday particulate matter concentration}
\end{center}
\begin{center}
Han Lin Shang\footnote{Postal address: Research School of Finance, Actuarial Studies and Statistics, Level 4, Building 26C, Kingsley Street, Acton ACT 2601, Australia; Telephone number: +61(2) 612 50535; Fax number: +61(2) 612 50087; Email address: hanlin.shang@anu.edu.au} \\
Australian National University
\end{center}
\vspace{.5in}

\begin{abstract}
Environmental data often take the form of a collection of curves observed sequentially over time. An example of this includes daily pollution measurement curves describing the concentration of a particulate matter in ambient air. These curves can be viewed as a time series of functions observed at equally spaced intervals over a dense grid. The nature of high-dimensional data poses challenges from a statistical aspect, due to the so-called `curse of dimensionality', but it also poses opportunities to analyze a rich source of information to better understand dynamic changes at short time intervals. Statistical methods are introduced and compared for forecasting one-day-ahead intraday concentrations of particulate matter; as new data are sequentially observed, dynamic updating methods are proposed to update point and interval forecasts to achieve better accuracy. These forecasting methods are validated through an empirical study of half-hourly concentrations of airborne particulate matter in Graz, Austria. 
\end{abstract}


\newpage

\section{Introduction}

Air pollution is composed of a mixture of compounds, including ozone, carbon monoxide, nitrogen oxides, sulfur dioxide, and particulate matter (PM). Increasing amount of epidemiological evidence implicates air pollution, particularly PM, as a major risk factor with important consequences to human health \citep{SHW10}. Therefore, government agencies constantly monitor the concentrations of PM. Accurate forecasts of PM will improve early warning systems, and are useful for public safety and implementing necessary policies. 

A number of statistical methods have been proposed to forecast PM \citep[see][for reviews]{DYK+09}. Among these, the popular models include neural networks \citep{PKK+11} and multiple linear modeling \citep{SHP08}. A comparison between these two methods is given in \cite{SKK06}. Recently, \cite{MMP+15} considered a mixture of linear regression models to forecast short-term PM concentration. A commonality in these works is that they often use discrete-time models, which depend on the time points at which the measurements are taken. Discrete-time models ignore the underlying dynamic changes of a continuous curve; that is, how the concentration of a PM shifts from a time point $t-1$ to $t$, where $t$ denotes a time variable that can be considered as an hour, a day or a month. If the concentrations of a PM are considered and analyzed as a discrete time series, we can not recover an underlying continuous stochastic process that generates these observations. When the stochastic process is smooth, one can also analyze the derivative information, such as by functional autoregressive regression \citep{MP07} and functional linear regression (FLR) \citep{MP09}. As pointed out by \cite{FGV02}, the rate of convergence of the kernel estimator of the regression function in a functional time series is superior to that obtained from a univariate time series. As an alternative, Functional Data Analysis techniques can often extract additional information contained in a time series of functions, such as their derivatives \citep[see for example,][Chapter 18]{RS06}. 

We study the daily curves of half-hourly concentrations of PM with an aerodynamic diameter of less than 10, abbreviated PM$_{10}$ hereafter \citep[see][]{SHP08, ANH15}. Using the stationary test proposed by \cite{HKR14}, we found that the daily pollution curves considered in this paper are stationary. Denote by $P_i(t_j), i = 1, \dots,n, j = 1,\dots,p$ as the concentration of PM$_{10}$ at time $t_j$ on day $i$. A square-root transformation was applied to the data to stabilize the variance, i.e., we work with the data $R_i(t_j) = \sqrt{P_i(t_j)}$. By interpolation, we consider the intraday curves as continuous curves constructed from
\begin{equation*}
\mathcal{X}_i(t) = R_i(t_j),\qquad  j = 1,\dots,p,
\end{equation*}
where $p$ denotes the total number of realized intraday measurement of PM$_{10}$. Once we have constructed a time series of functions, we will work directly with a continuous functional time series. In the intraday PM$_{10}$ data described in Section~\ref{sec:2}, there are $p=48$ half-hourly time intervals representing 24 hours of intraday measurements.

In the statistical literature, there has been a large amount of research on the development of functional time series forecasting methods \citep[see][for a general background]{HS09, HK10, HK12}. From a parametric viewpoint, \cite{Bosq91, Bosq00} proposed the functional autoregressive (FAR) of order 1, and derived one-step-ahead forecasts that are based on a regularized form of the Yule-Walker equations. \cite{KK16} proposed the functional moving average (FMA) process and introduced an innovations algorithm to obtain the best linear predictor. \cite{KKW16} proposed the FARMA process where a dimension reduction technique was used to reduce infinite-dimensional objects to a finite dimension, and thus vector autoregressive (VAR) models can be deployed. From a nonparametric viewpoint, \cite{BCS00} proposed functional kernel regression to measure the temporal dependence by a similarity measure characterized by the notions of neighborhood distance and bandwidth. From a semi-parametric viewpoint, \cite{AV08} put forward a semi-functional partial linear model that combines parametric and nonparametric models, and this semi-functional partial linear model allows us to consider additional covariates and to use a continuous path in the past to predict the future values of the stochastic process.

Among many modeling techniques, functional principal component analysis (FPCA) has been used extensively for dimension reduction for a functional time series. As a data-driven basis function decomposition, FPCA can collapse an infinite-dimensional object to a finite dimension, without losing much information. We adopt the approach by \cite{HS09} and \cite{ANH15}, who applied a functional principal component regression to decompose a time series of functions into a set of functional principal components and their associated scores. The temporal dependence in the original functional time series is inherited by the correlation within each set of principal component scores and the possible correlation among principal component scores. While \cite{HS09} applied a univariate time series forecasting technique (autoregressive integrated moving average (ARIMA)), to forecast these scores, \cite{ANH15} considered a multivariate time series forecasting technique (a VAR model). Conditioning on the historical curves and estimated functional principal components, the point forecasts are obtained by multiplying the forecast principal component scores by the estimated functional principal components. Since this method uses either univariate or multivariate time series forecasts, we call it the ``TS method", described in Section~\ref{sec:3}. 

When functional time series are segments of a univariate time series, the most recent trajectory is observed sequentially, such as the intraday PM$_{10}$ data described in Section~\ref{sec:2}. These intraday data present a source of information highlighting the PM$_{10}$ changes happening during a day. By incorporating this new information, we can update our point and interval forecasts in the hope of achieving better forecast accuracy \citep[see also][]{SH08, Shen09, SH11}. We present two new dynamic updating methods for updating point forecasts in Section~\ref{sec:4}, and propose a nonparametric bootstrap method for updating interval forecasts in Section~\ref{sec:5}. Using the forecast error measures given in Section~\ref{sec:6}, we examine the point and interval forecast accuracies in Section~\ref{sec:7}. Our conclusions are given in Section~\ref{sec:8}, along with some reflections on how the methods presented here may be further extended. 

\section{Forecasting methods}\label{sec:3}

\subsection{Functional principal component regression}

Let $(\mathcal{X}_i: i\in Z)$ be an arbitrary stationary functional time series. It is assumed that the observations $\mathcal{X}_i$ are elements of the Hilbert space $H=L^2(\mathcal{I})$ equipped with the inner product $\langle x,y\rangle = \int_{\mathcal{I}} x(t)y(t)dt$ and $t$ represents a continuum within a function support range denoted by $\mathcal{I}$. Each function is a square integrable function satisfying $\|\mathcal{X}_i\|^2 = \int_{\mathcal{I}}\mathcal{X}_i^2(t)dt <\infty$. All random functions are defined on a common probability space $(\Omega, A, P)$. The notation $\mathcal{X}\in L^{\rho}_H(\Omega, A, P)$ is used to indicate for some $\rho>0$, $\text{E}\left(\|\mathcal{X}\|^{\rho}\right)<\infty$. When $\rho=1$, $\mathcal{X}(t)$ has the mean curve $\mu(t) = \text{E}\left[\mathcal{X}(t)\right]$; when $\rho=2$, the covariance operator $\mathcal{K}(s, t) = \text{Cov}[\mathcal{X}(s),\mathcal{X}(t)] = \text{E}\left\{[\mathcal{X}(s) - \mu(s)][\mathcal{X}(t) - \mu(t)]\right\}$ is defined by
\begin{equation*}
\mathcal{K}(s,t) = \sum^{\infty}_{k=1}\lambda_k\phi_k(s)\phi_k(t), \qquad s, t \in \mathcal{I}, 
\end{equation*}
where $\phi_k(t)$ denotes the $k^{\text{th}}$ orthonormal principal component, and $\lambda_k$ denotes the $k^{\text{th}}$ eigenvalue \citep[see][]{Karhunen46,Loeve46}. In the intraday PM$_{10}$ curves, $t\in (0, 24]$ is the function support range, where $t_1$ symbolizes the beginning point (just after midnight) and $t_{48}$ symbolizes the end point (midnight).

Based on the separability of the Hilbert space, the Karhunen-Lo\`{e}ve expansion of a stochastic process $\mathcal{X}$ can be expressed as
\begin{equation}
\mathcal{X}(t) = \mu(t) + \sum^{\infty}_{k=1}\beta_k\phi_k(t), \label{eq:3.1}
\end{equation}
where the principal component scores $\beta_k$ are given by the projection of $\left[\mathcal{X}(t) - \mu(t)\right]$ in the direction of the $k^{\text{th}}$ eigenfunction $\phi_k$, i.e., $\beta_k = \langle\mathcal{X} - \mu, \phi_k\rangle$.

Expansion~\eqref{eq:3.1} facilitates dimension reduction as the first $K$ terms often provide a good approximation to the infinite sums and thus the information contained in $\mathcal{X}(t)$ can be adequately summarized by the $K$-dimensional vector $(\beta_1,\dots,\beta_K)$. The approximated processes can be defined as
\begin{equation}
\mathcal{X}(t) = \mu(t) + \sum^K_{k=1}\beta_k\phi_k(t)+e(t), \label{eq:fpca}
\end{equation}
where $K$ represents the number of retained principal components, and $e(t)$ represents the error term, containing the functional principal components excluded from the first $K$ terms. Although this could be a research topic on its own, there are several approaches for selecting $K$: 
\begin{inparaenum}[(1)]
\item scree plots or the fraction of variance explained by the first few functional principal components \citep{Chiou12};
\item pseudo-versions of the Akaike information criterion and Bayesian information criterion \citep{Yao05};
\item cross-validation with one-curve-leave-out \citep{RS91}; or
\item the bootstrap technique \citep{HV06}.
\end{inparaenum}
Here, the value of $K$ is chosen as the minimum that reaches a certain level of the proportion of the total variance explained by the $K$ leading components such that
\begin{equation*}
K = \argmin_{K: K\geq 1}\left\{\sum^K_{k=1}\widehat{\lambda}_k\Big/\sum^{\infty}_{k=1}\widehat{\lambda}_k \mathds{1}\{\widehat{\lambda}_k>0\}\geq \delta\right\},
\end{equation*}
where $\mathds{1}\{\widehat{\lambda}_k>0\}$ is to exclude possible zero eigenvalues, and $\mathds{1}\{\cdot\}$ represents the binary indicator function.

In practice, we reconstruct a time series of functions $\bm{\mathcal{X}}(t) = \{\mathcal{X}_1(t),\dots,\mathcal{X}_n(t)\}$, from which the mean and covariance functions can be empirically estimated. From the empirical covariance function, we extract the empirical functional principal component functions $\bm{\widehat{\Phi}}(t) = \left[\widehat{\phi}_1(t),\dots,\widehat{\phi}_K(t)\right]$. Conditional on the mean function $\widehat{\mu}(t)$, the estimated functional principal components $\bm{\widehat{\Phi}}(t)$ and a time series of historical functions $\bm{\mathcal{X}}(t)$, the point forecasts of $\mathcal{X}_{n+h}(t)$ can be obtained as
\begin{equation*}
\widehat{\mathcal{X}}_{n+h|n}(t) = \text{E}[\mathcal{X}_{n+h}(t)|\widehat{\mu}(t), \bm{\widehat{\Phi}}(t),\bm{\mathcal{X}}(t)]= \widehat{\mu}(t) + \sum^K_{k=1}\widehat{\beta}_{n+h|n,k}\widehat{\phi}_k(t), 
\end{equation*}
where $\widehat{\mu}(t) =  \frac{1}{n}\sum^n_{i=1}\mathcal{X}_i(t)$, $\widehat{\phi}_k(t)$ represents the $k^{\text{th}}$ estimated functional principal component, and $\widehat{\beta}_{n+h|n,k}$ represents point forecasts of $\beta_{n+h,k}=\langle \mathcal{X}_{n+h} - \mu, \phi_k \rangle$ using a univariate or multivariate time series forecasting technique (see Section~\ref{sec:3.3} for more details). 

\subsection{Robust functional principal component analysis}

In the presence of outliers, the covariance operator $\mathcal{K}(s,t) = \text{cov}[\mathcal{X}(s),\mathcal{X}(t)]$ may not be robust against outliers. As a result, the estimated functional principal components extracted from the covariance operator can be erroneous and this could lead to inferior estimation and forecasting accuracies. To bypass this issue, we consider a robust FPCA, namely the two-step algorithm of \cite{HU07}. The robust functional principal components are extracted by down-weighting the effect of outliers. This procedure begins with a robust FPCA algorithm, such as that proposed by \cite{HRV02}, then calculates the integrated squared error of curve $i$ as 
\begin{equation*}
v_i = \int_t \Big[\mathcal{X}_i(t) - \sum^K_{k=1}\beta_{i,k}\phi_k(t)\Big]^2dt,\qquad \text{for} \quad  i=1,\dots,n,
\end{equation*}
where outlying curves tend to have large values of $v_i$. A set of weights $w_i$ is then assigned if $v_i<s+\lambda\sqrt{s}$, where $\lambda>0$ represents a tuning parameter for controlling the amount of robustness and $s$ is the median of $\{v_1,\dots,v_n\}$. As described in \cite{HU07}, the efficiency of this procedure follows a cumulative normal distribution; when $\lambda=2.33$, the efficiency is $\Phi\left(2.33/\sqrt{2}\right)=95\%$, which implies that 5\% of observations are treated as outliers.

\subsection{Univariate and multivariate time series forecasting techniques}\label{sec:3.3}

To obtain $\widehat{\beta}_{n+h|n,k}$, \cite{HS09} considered a univariate time series forecasting technique; namely, an ARIMA model. This univariate time series forecasting technique is able to model non-stationary time series containing a stochastic trend component. As the intraday PM$_{10}$ curves include only 182 days and thus do not contain seasonality, the ARIMA$_k$ for each observation of the $k^{\text{th}}$ principal component scores has the general form of
\begin{equation*}
\left(1-\varphi_1^{(k)} B-\cdots - \varphi_p^{(k)}B^{p^{(k)}}\right)\left(1-B\right)^{d^{(k)}}\beta_{i,k} = \alpha + \left(1+\theta_1^{(k)} B + \cdots+\theta_q^{(k)} B^{q^{(k)}}\right)w_{i,k},
\end{equation*}
where $\alpha$ represents the intercept, $(\varphi_1^{(k)},\cdots,\varphi_p^{(k)})$ denote the coefficients associated with autoregressive component, $(\theta_1^{(k)},\dots,\theta_q^{(k)})$ denote the coefficients associated with moving average component, $B$ denotes the backshift operator, $d^{(k)}$ denotes the differencing operator for $k^{\text{th}}$ principal component scores, $\beta_{i,k}$ represents the $k^{\text{th}}$ estimated principal component scores for the $i^{\text{th}}$ observation and $w_{i,k}$ represents its error term. The optimal ARIMA$_k$ model is selected based on an information criterion, and then the parameters can be estimated by the maximum likelihood estimator. Computationally, the automatic ARIMA algorithm of \cite{HK08} has been implemented for selecting optimal orders based on an information criterion, such as the corrected Akaike information criterion, which is particularly suitable for small sample size.  

Depending on the structure of the data, univariate time series forecasting techniques can be quick and efficient in some cases, but less accurate in others \citep{Tsay13, ANH15}. Although the functional principal component score vectors have no instantaneous correlation, this does not imply that auto-covariances at lags greater than zero remain diagonal. Hence, the univariate time series modeling may result a loss of valuable information hidden in the dependence of the principal component score matrix. To rectify this problem, \cite{ANH15} suggested to using a multivariate time series model, and the most commonly used multivariate time series model is the VAR model because 
\begin{inparaenum}[(1)]
\item this model is relatively easy to estimate using ordinary least squares (OLS), maximum likelihood or Bayesian method; 
\item the properties of the VAR model have been studied extensively \citep[see, e.g.,][]{Tsay13};
\item VAR models can be viewed as a multivariate multiple linear regression. 
\end{inparaenum}

The multivariate time series of principal component scores $\bm{\beta}_k = [\beta_{1,k},\dots,\beta_{n,k}]^{\top}$ follows a VAR model of order $\vartheta$, VAR($\vartheta$) if
\begin{equation*}
\bm{\beta}_k = \bm{\phi}_0 + \sum^{\vartheta}_{\upsilon=1}\bm{\phi}_{\upsilon}\bm{\beta}_{k-\upsilon}+\bm{a}_k,
\end{equation*}
where $\bm{\phi}_0$ is a $n$-dimensional constant vector, $\bm{\phi}_{\upsilon}$ are $n\times n$ matrices for $\upsilon>0$ and $\bm{\phi}_{\upsilon} \neq 0$, $\bm{a}_k$ is a set of independent and identically distributed (iid) random error vectors with a mean of zero and a positive-definite covariance matrix that has all positive eigenvalues, and $\vartheta$ can be determined by an information criterion, such as the Akaike information criterion. 

Via a multivariate linear regression model, the VAR$(\vartheta)$ can be re-written as VAR(1),
\begin{equation*}
\bm{\beta}_k = \bm{B}^{\top}\bm{x}_k + \bm{a}_k,\qquad k = \vartheta+1, \dots, n
\end{equation*}
with $\bm{x}_k = \left(1, \bm{\beta}_{k-1}^{\top},\dots,\bm{\beta}_{k-\vartheta}^{\top}\right)^{\top}$ and $\bm{B} = \left(\bm{\phi}_0^{\top}, \bm{\phi}_1^{\top}, \dots, \bm{\phi}_{\vartheta}\right)^{\top}$ contains all unknown regression coefficients. Let $\bm{X} = \left(\bm{x}_{\vartheta+1},\dots,\bm{x}_n\right)^{\top}$ denote the matrix containing the values of the explanatory variables and $\bm{Y} = \left(\bm{\beta}_{\vartheta+1},\dots,\bm{\beta}_n\right)^{\top}$ be the matrix of response. The unknown regression coefficients can then be estimated by OLS as
\begin{equation*}
\widehat{\bm{B}}_{\text{OLS}} = \left(\bm{X}^{\top}\bm{X}\right)^{-1}\bm{X}^{\top}\bm{Y}.
\end{equation*}
Conditioning on the estimated principal components and past curves, the $h$-step-ahead forecast principal component scores and forecast curve are
\begin{align*}
\widehat{\beta}_{n+h|n} &= \widehat{\bm{B}}_{\text{OLS}}^{\top}\times \bm{x}_{n+h}, \\
\widehat{\mathcal{X}}_{n+h|n}(t) &= \widehat{\beta}_{n+h|n}\times \widehat{\bm{\phi}}(t),
\end{align*}
where $\widehat{\bm{\phi}}=\left[\widehat{\phi}_1(t),\dots,\widehat{\phi}_K(t)\right]$ and $\widehat{\bm{\beta}}_{n+h|n} = \big[\widehat{\beta}_{n+h|n,1},\dots,\widehat{\beta}_{n+h|n,K}\big]$.

As noted by \cite{PS07}, if the series (i.e., principal component score vectors) are very weakly related, there is a marginal improvement on the univariate forecasts by using the joint dynamics of the series. Unlike univariate time series forecasting techniques, a VAR model is restricted to stationary multivariate time series. Although a vector error correction model can deal with non-stationary multivariate time series, the number of parameters remains comparably larger than those from the univariate time series model, and this may hinder the accuracy of parameter estimation, especially for a small sample size. On the other hand, when the present value of one of the time series depends strongly on the past values of the others, there are clear advantages in the multivariate time series forecasts in comparison to univariate time series forecasts. Through the empirical data analyses in Section~\ref{sec:7}, we compare the point and interval forecast accuracies between the univariate and multivariate time series forecasting techniques.

\section{Updating point forecasts}\label{sec:4}

When a functional time series is constructed from segments of a longer univariate time series, the most recent curve is observed sequentially and thus may not be a complete curve. When we have observed the first $m_0$ time periods of $\mathcal{X}_{n+1}(t)$, denoted by $\mathcal{X}_{n+1}(t_e) = [\mathcal{X}_{n+1}(t_1),\dots,\mathcal{X}_{n+1}(t_{m_0})]^{\top}$, we are particularly interested in forecasting the data in the remainder of day $n+1$, denoted by $\mathcal{X}_{n+1}^l(\mathrm{t})$ where $\mathrm{t}\in \mathcal{I}_l$ and $\mathcal{I}_l\in (m_0, p]$ represents a function support range for the remaining data period (note that $\mathcal{X}_{n+1}(t_e)$ represents a set of observed data points in function $\mathcal{X}_{n+1}$, not function $\mathcal{X}_{n+1}$ evaluated at $t_e$). However, the TS method presented in Section~\ref{sec:3} does not utilize the most recent data, i.e., the partially observed curve. The TS forecasts of $\mathcal{X}_{n+1}^l(\mathrm{t})$ is given by
\begin{equation}
\widehat{\mathcal{X}}_{n+1|n}^{l,\text{TS}}(\mathrm{t}) = \widehat{\mu}^l(\mathrm{t}) + \sum^K_{k=1}\widehat{\beta}_{n+1|n,k}^{\text{TS}}\widehat{\phi}_k^l(\mathrm{t}),\label{eq:TS}
\end{equation}
where $\widehat{\mu}^l(\mathrm{t})$ represents the mean curve for the remaining time period, $\widehat{\beta}_{n+1|n,k}^{\text{TS}}$ denotes the univariate or multivariate time series forecasts of the principal component scores described in Section~\ref{sec:3.3} and these forecasts form the essence for the block moving (BM) method in Section~\ref{sec:4.1} and FLR in Section~\ref{sec:4.2}, and $\widehat{\phi}_k^l(\mathrm{t})$ represents the $k^{\text{th}}$ functional principal component corresponding to the remaining time period. 

In order to improve point forecast accuracy, it is desirable to dynamically update the point forecasts for the remaining time period of the most recent curve $n+1$ by incorporating the newly arrived information. To address this issue, \cite{SH08, Shen09} and \cite{SH11} have proposed several dynamic updating methods, but they are mainly discrete-time models. In what follows, we shall introduce two new dynamic updating methods to achieve better forecast accuracy.

\subsection{Block moving (BM) method}\label{sec:4.1}

The BM method uses the TS method, but it redefines the beginning and end time points of our curves. Because time is a continuous variable, we can change the function support from $(0,24]$ to $(0,m_0]\cup (m_0,24]$. 

The redefined data are shown in Figure~\ref{fig:N1}, where the bottom box has moved to become the top box. The colored region shows the data loss in the first year. The partially observed last trajectory under the old function support completes the last trajectory under the new function support. 

\tikzset{decorate with/.style={fill=cyan!20,draw=cyan}}
\begin{figure}[!htbp]
\begin{center}
\scalebox{.56}{
\begin{tikzpicture}
\draw (0,0) node[anchor=north east]{$t_p$} rectangle (12,8);
\draw (1.1,4.1) rectangle (13.1,12);
\draw (0, 8) node[anchor=north east]{$t_1$} -- (0, 0);
\draw (0,4.1) node[anchor=north east]{  \begin{rotate}{90} \hspace {-.5in} Dimensionality \end{rotate} \hspace{.05in} $t_{m_0}$} -- (1.1, 4.1);
\draw[dashed] (1.1,0) -- (1.1,4.1);
\draw[dashed] (2.2,12) -- (2.2,8);
\draw[dashed] (14.2, 12) -- (14.2, 8);
\draw (11.86, 8) node[anchor=south west]{$n+1$} -- (13, 8);
\draw[dashed] (13, 8) -- (14.2, 8);
\draw[dashed] (13, 12) -- (14.2, 12);
\draw[dashed] (13.1, 4.1) -- (13.1, 0);
\draw[dashed] (12, 0) -- (13.1, 0);
\draw (8.5,0) node[anchor=north east]{Number of curves} -- (12,0);
\draw(0.4,8) node[anchor=south west]{1} -- (1.1,8);
\draw [decorate with=rectangle] (0.0305,4.1365) -- (0.0305,7.98) --  (1.074,7.98) -- (1.074,4.1365);
\end{tikzpicture}}
\end{center}
\caption{Dynamic updating via the BM approach. The colored region shows the data loss in the first year. The forecasts for the rest of year $n+1$ can be updated by the TS forecasts applied to the top block.}\label{fig:N1}
\end{figure}
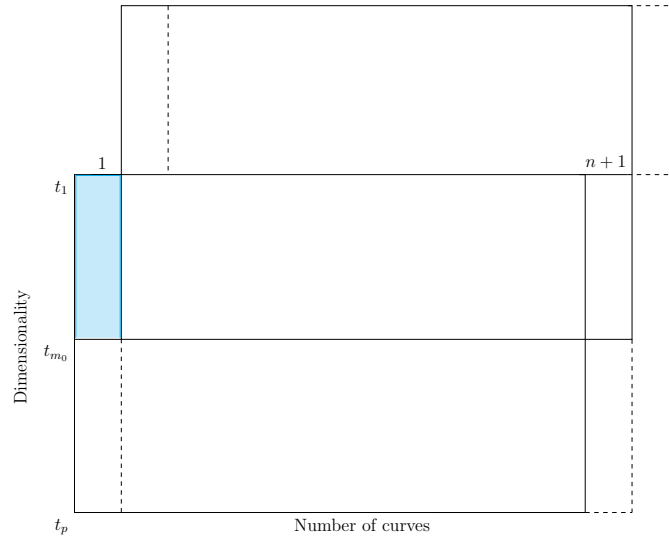

As a result, a partially observed curve can be completed at a loss of some data in the first curve. When the number of curves is large, the data loss in the first curve will have minimal effect on the forecasts, as the forecasts are not greatly depend on the observations from the distant past. The updated forecasts can be obtained by the TS method described in~\eqref{eq:TS} with univariate and multivariate time series forecasting techniques. 

\subsection{Functional linear regression}\label{sec:4.2}

As an alternative to time series techniques, we consider the FLR to update forecasts \citep[see also][]{MSS11, Chiou12}. The FLR is given by
\begin{equation}
\mathcal{X}_{n+1}^l(\mathrm{t}) = \mu^l(\mathrm{t})+\int_{\mathrm{s}\in \mathcal{I}_e} \left[\mathcal{X}_{n+1}^e(\mathrm{s})-\mu^e(\mathrm{s})\right]\tau(\mathrm{s},\mathrm{t})d\mathrm{s}+e^l_{n+1}(\mathrm{t}),\qquad \mathrm{s}\in \mathcal{I}_e, \quad \mathrm{t}\in \mathcal{I}_l,  \label{eq:flr}
\end{equation}
where $\mathcal{I}_e\in [1,m_0]$ and $\mathcal{I}_l\in (m_0,p]$ represent two function support ranges for the partially observed data and remaining data periods; $\mu^e(\mathrm{s})$ and $\mu^l(\mathrm{t})$ represent two mean functions for the partially observed data and remaining data periods; $\mathcal{X}_{n+1}^e(\mathrm{s})$ and $\mathcal{X}_{n+1}^l(\mathrm{t})$ represent functional predictor and functional response variables. Equation~\eqref{eq:flr} can be viewed as function-on-function linear regression \citep[see also][Chapter 16]{RS06}, where $\tau(\mathrm{s},\mathrm{t})$ and $e^l_{n+1}(\mathrm{t})$ denote the regression coefficient function and error function, respectively.

For estimating $\tau(\mathrm{s},\mathrm{t})$, we project a time series of functions onto functional principal component scores. Through FPCA, we obtain

\begin{minipage}{7cm}
\begin{align}
\mathcal{X}_i^e(\mathrm{s})&=\mu^e(\mathrm{s})+\sum^{\infty}_{k=1}\xi_{i,k}\phi_k^e(\mathrm{s}), \notag\\
&=\mu^e(\mathrm{s}) + \sum^K_{k=1}\xi_{i,k}\phi_k^e(\mathrm{s})+\eta_i^e(\mathrm{s}), \notag
\end{align}
\end{minipage}
\hfill
\begin{minipage}{7cm}
\begin{align}
\mathcal{X}_i^l(\mathrm{t}) &= \mu^l(\mathrm{t}) + \sum^{\infty}_{m=1}\zeta_{i,m}\psi_m^l(\mathrm{t}), \label{eq:flr_after}\\
&=\mu^l(\mathrm{t})+\sum^{M}_{m=1}\zeta_{i,m}\psi_m^l(\mathrm{t})+v_i^l(\mathrm{t}), \notag
\end{align}
\end{minipage}
\vspace{.2in}

\noindent where $\phi_k^e(\mathrm{s})$ and $\psi_m^l(\mathrm{t})$ denote the $k^{\text{th}}$ and $m^{\text{th}}$ functional principal components associated with the partially observed and remaining data periods; $\xi_{i,k}$ and $\zeta_{i,m}$ are the principal component scores of $\mathcal{X}_i^e(\mathrm{s})$ and $\mathcal{X}_i^l(\mathrm{t})$; $K$ and $M$ are retained numbers of components; $\eta_i^e(\mathrm{s})$ and $v_i^l(\mathrm{t})$ represent the error functions associated with the partially observed and remaining data periods, due to model truncations. 

Let $\bm{\zeta}_m=[\zeta_{1,m},\dots,\zeta_{n,m}]^{\top}$ and $\bm{\xi}_k=[\xi_{1,k},\dots,\xi_{n,k}]^{\top}$. By treating $\bm{\zeta} = [\bm{\zeta}_1,\dots,\bm{\zeta}_M]$ as a response variable and $\bm{\xi} = [\bm{\xi}_1,\dots,\bm{\xi}_K]$ as an explanatory variable, their relationship can be expressed as
\begin{equation}
\bm{\zeta} = \bm{\xi} \times \bm{\varsigma},\label{eq:ols_flr}
\end{equation}
where $\bm{\varsigma}$ can be estimated by OLS, given by
\begin{equation}
\widehat{\bm{\varsigma}} =\left(\bm{\xi}^{\top}\bm{\xi}\right)^{-1} \bm{\xi}^{\top}\bm{\zeta},\label{eq:varsigma}
\end{equation}
where $\bm{\xi}\times \bm{\zeta}$ can be modeled jointly by their cross-covariance structure
\begin{equation*}
\int_{\mathrm{t}\in \mathcal{I}_l}\int_{\mathrm{s}\in \mathcal{I}_e} \phi_k(\mathrm{s})\psi_m(\mathrm{t})\text{cov}\left[\bm{\mathcal{X}}^e(\mathrm{s}),\bm{\mathcal{X}}^l(\mathrm{t})\right]dsdt,\qquad k=1,\dots,K, \quad m = 1,\dots,M,
\end{equation*}
where $\bm{\mathcal{X}}^e(\mathrm{s}) = \left[\mathcal{X}_1^e(\mathrm{s}),\dots,\mathcal{X}_n^e(\mathrm{s})\right]^{\top}$ and $\bm{\mathcal{X}}^l(\mathrm{t}) = \left[\mathcal{X}_1^l(\mathrm{t}),\dots,\mathcal{X}_n^l(\mathrm{t})\right]^{\top}$ be two vectors of functions corresponding to the partially observed data and remaining data periods.

The point forecast of $\mathcal{X}_{n+1}^l(\mathrm{t})$ can be obtained from~\eqref{eq:flr_after} and expressed as
\begin{equation}
\widehat{\mathcal{X}}_{n+1}^l(\mathrm{t}) = \widehat{\mu}^l(\mathrm{t}) + \sum^{\infty}_{m=1}\zeta_{n+1,m}\psi_m^l(\mathrm{t}). \label{eq:flr_point}
\end{equation}
From~\eqref{eq:ols_flr},~\eqref{eq:flr_point} can be approximated as 
\begin{equation*}
\widehat{\mathcal{X}}_{n+1}^l(\mathrm{t}) \approx  \widehat{\mu}^l(\mathrm{t})+\widehat{\bm{\xi}}_{n+1}\times \widehat{\bm{\varsigma}} \times\widehat{\bm{\psi}}^l(\mathrm{t}),
\end{equation*}
where $\widehat{\bm{\varsigma}}$ is estimated from~\eqref{eq:varsigma}, and $\widehat{\bm{\psi}}^l(\mathrm{t}) = \big[\widehat{\psi}_1^l(\mathrm{t}), \dots, \widehat{\psi}_M^l(\mathrm{t})\big]^{\top}$.

\section{Interval forecast methods}\label{sec:5}

Prediction intervals are a valuable tool for measuring the probabilistic uncertainty associated with point forecasts. As emphasized in \cite{Chatfield93}, it is important to provide interval forecasts as well as point forecasts so as to 
\begin{inparaenum}[(1)]
\item assess future uncertainty;
\item enable different strategies to be planned for a range of possible outcomes indicated by the interval forecasts;
\item compare forecasts from different methods more thoroughly;
\item and explore different scenarios based on different assumptions.
\end{inparaenum}

To quantify forecast uncertainty, it is essential to understand the sources of errors. In our functional principal component regression, two sources of errors are from estimating the regression coefficient function and model errors. In Section~\ref{sec:5.1}, we describe a nonparametric bootstrap method for constructing one-step-ahead prediction intervals for the TS method. In Section~\ref{sec:5.2}, we show how the prediction intervals can be updated through the BM method and FLR. 

\subsection{Nonparametric prediction interval}\label{sec:5.1}

Since our focus is on short-term time series forecasting, we can measure the one-step-ahead forecast errors for estimated principal component scores, given by
\begin{equation*}
\widehat{\omega}_{j,k} = \widehat{\beta}_{n-j+1,k}-\widehat{\beta}_{n-j+1|n-j,k}\qquad \text{for}\quad j=1,\dots,n-K,
\end{equation*}
where $\widehat{\beta}_{n-j+1,k}$ denotes the $k^{\text{th}}$ estimated principal component score for day $n-j+1$; $\widehat{\beta}_{n-j+1|n-j,k}$ denotes its one-step-ahead forecast obtained by either a univariate or multivariate time series technique; $K$ represents the number of retained principal components in~\eqref{eq:fpca}; and $n-K$ can be viewed as the total number of in-sample principal component score errors. These one-step-ahead forecast errors can then be sampled with replacement to produce a bootstrap sample of $\beta_{n+1,k}$:
\begin{equation*}
\widehat{\beta}_{n+1|n,k}^b = \widehat{\beta}_{n+1|n,k} + \widehat{\omega}_{*,k}^b,\qquad \text{for}\quad b = 1,\dots,B,
\end{equation*}
where $\widehat{\omega}_{*,k}^b$ denotes the bootstrapped forecast errors by sampling with replacement from $(\widehat{\omega}_{1,k},\dots,$
$\widehat{\omega}_{n-K,k})$, and $B$ is the number of bootstrap replications. 

When the first $K$ functional principal component decomposition in~\eqref{eq:fpca} approximate the data relatively well, the model residuals should be iid random noise. Therefore, we can bootstrap these model residual function $\widehat{e}_{n+1}(t)$ by sampling with replacement from the historical residual functions $\left\{\widehat{e}_2(t),\dots,\widehat{e}_n(t)\right\}$. 

Adding these two sources of errors, we obtain $B$ bootstrapped forecasts of $\mathcal{X}_{n+1}(t)$, given by
\begin{equation*}
\widehat{\mathcal{X}}_{n+1|n}^b(t) = \widehat{\mu}(t) + \sum^K_{k=1}\widehat{\beta}_{n+1|n,k}^b\widehat{\phi}_k(t) + \widehat{e}_{n+1}^b(t).
\end{equation*}
Hence, the $100(1-\alpha)\%$ pointwise prediction intervals are defined as $\alpha/2$ and $(1-\alpha/2)$ empirical quantiles of $\left\{\widehat{\mathcal{X}}_{n+1|n}^1(t_j), \dots,\widehat{\mathcal{X}}_{n+1|n}^B(t_j),\quad j=1,\dots,p\right\}$. This nonparametric prediction interval approach will also work for the BM method, with a modification of the function support range.

\subsection{Updating the prediction interval}\label{sec:5.2}

We can also construct prediction intervals by FLR through bootstrapping. With the FLR in~\eqref{eq:flr}, the bootstrapped forecasts $\mathcal{X}^{l,b}_{n+1}(\mathrm{t})$ can be obtained as
\begin{equation*}
\widehat{\mathcal{X}}^{l,b}_{n+1}(\mathrm{t}) = \mu^l(\mathrm{t}) + \int_{\mathrm{s}\in\mathcal{I}_e} \left[\mathcal{X}_{n+1}^e(\mathrm{s})-\mu(\mathrm{s})\right]\widehat{\tau}^b(\mathrm{s},\mathrm{t})d\mathrm{s} + \widehat{e}^{l,b}_{n+1}(\mathrm{t}),
\end{equation*}
where $\widehat{\tau}^b(\mathrm{s},\mathrm{t})$ represents the bootstrapped regression coefficient estimates, and $\widehat{e}_{n+1}^{l,b}(\mathrm{t})$ represents the bootstrapped error term associated with the remaining time period. The $\widehat{\tau}^b(\mathrm{s},\mathrm{t})$ captures the parameter variability in the estimation of regression coefficient function, while the $\widehat{e}^{l,b}_{n+1}(\mathrm{t})$ measures the model variability. 

Assuming the one-step-ahead forecast errors do not correlate, we implement the iid bootstrap method by sampling with replacement from historical errors $\left\{\widehat{e}^{l}_2(\mathrm{t}),\dots,\widehat{e}^l_n(\mathrm{t})\right\}$. The bootstrapped $\widehat{\tau}^b(\mathrm{s},\mathrm{t})$ can then be obtained by bootstrapping the original functional time series via functional principal component decomposition, expressed as
\begin{equation*}
\widehat{\mathcal{X}}_i^b(t) = \widehat{\mu}(t) + \sum^{\infty}_{k=1}\widehat{\beta}^b_{i,k}\widehat{\phi}_k(t), \qquad i=1,\dots,n,
\end{equation*}
where $\left(\widehat{\beta}^b_{1,k},\dots,\widehat{\beta}^b_{n,k}\right)$ represents the bootstrapped $k^{\text{th}}$ estimated principal component scores.  With a set of bootstrapped data $\{\widehat{\mathcal{X}}_1^b(t),\dots,\widehat{\mathcal{X}}_n^b(t)\}$, we apply the FLR in~\eqref{eq:flr} to obtain bootstrapped estimates of regression coefficient function, $\widehat{\tau}^b(s,t)$. 

Among many bootstrap techniques for multivariate time series $\left\{\bm{\beta}_1,\dots,\bm{\beta}_K\right\}$, we use the maximum entropy bootstrap method proposed by \cite{Vinod04}. The advantages of the maximum entropy bootstrap technique or time series are: 
\begin{inparaenum}[(1)]
\item stationarity of principal component scores is not required;
\item the bootstrap technique computes the ranks of a time series, thus it is robust;
\item bootstrap samples satisfy the ergodic theorem, central limit theorem and mean preserving constraint;
\item bootstrap samples are adjusted so that the population variance of the ME density equals that of the original data.
\end{inparaenum}
An algorithm of the maximum entropy bootstrap is described in \cite{VD09}, and it is implemented in the \verb meboot.pdata.frame \ function of the \textit{meboot} package in \textsf{R} \citep{Team15}.

\subsection{Prediction bands}\label{sec:band}

We also consider the construction of uniform prediction intervals. The aim is to find parameters $\underline{\xi}_{\alpha}$ and $\overline{\xi}_{\alpha}\geq 0$ such that for a given $\alpha\in (0,1)$ and standard deviation function $\gamma: [0,24)\rightarrow [0,\infty)$, the empirical coverage probability is as close as the nominal coverage probability.
\begin{align*}
P\left(\widehat{\mathcal{X}}_{n+1|n}(t) - \underline{\xi}_{\alpha}\gamma(t) \leq \mathcal{X}_{n+1}(t) \leq \widehat{\mathcal{X}}_{n+1|n}(t) + \overline{\xi}_{\alpha}\gamma(t), \quad \forall t\in [0, 24)\right) &= \alpha, \\
P\left(- \underline{\xi}_{\alpha}\gamma(t) \leq \widehat{\varepsilon}_{n+1}(t) \leq \overline{\xi}_{\alpha}\gamma(t), \quad \forall t\in[0,24)\right) &= \alpha.
\end{align*}
Since $\widehat{\varepsilon}_{n+1}(t)$ is not observable, it can be estimated via bootstrapping from the observed residuals $\left\{\widehat{\varepsilon}_{K+1}(t), \dots, \widehat{\varepsilon}_n(t)\right\}$, where 
$\gamma(t)=\text{sd}\left\{\widehat{\varepsilon}_{K+1}(t), \dots, \widehat{\varepsilon}_n(t)\right\}$ denotes the standard deviation. The residuals $\left(\widehat{\varepsilon}_{K+1}(t),\dots,\widehat{\varepsilon}_n(t)\right)$ are then expected to be approximately stationary and by a law of large numbers effect, to satisfy
\begin{equation*}
\frac{1}{n-(K+1)}\sum^{n}_{k=K+1}I\left(-\underline{\xi}_{\alpha}\gamma(t) \leq \widehat{\varepsilon}_k(t) \leq \overline{\xi}_{\alpha}\gamma(t)\right)\approx P\left[-\underline{\xi}_{\alpha}\gamma(t) \leq \mathcal{X}_{n+1}(t) - \widehat{\mathcal{X}}_{n+1}(t)\leq \overline{\xi}_{\alpha}\gamma(t)\right].
\end{equation*}
Typically the constants $\underline{\xi}_{\alpha}$ and $\overline{\xi}_{\alpha}$ are chosen to be equal, and the optimal value can be determined through an optimization algorithm, such as the \verb optim \ function in R.

\section{Intraday PM$_{10}$ curves}\label{sec:2}

Let $\{Z_w, w\in [1, N]\}$ be a seasonal univariate time series, which has been observed at $N$ equispaced times. When the seasonal pattern is strong, one way to model the time series nonparametrically is to use ideas from functional data analysis. We divide the observed time series into $n$ trajectories, and then consider each trajectory of length $p$ as a curve rather than as $p$ distinct points. The functional time series is then given by
\begin{equation*}
\mathcal{X}_i(t_j) = \{Z_w,\quad w\in (p(i-1), pi]\}, \qquad i=1,\dots,n,
\end{equation*}
where $p=48$ and $0<t_1\leq t_2 \leq \cdots \leq t_{48}=24$.

\begin{figure}[!htbp]
\centering
\subfloat[A univariate time series display]
{\includegraphics[width=8.5cm]{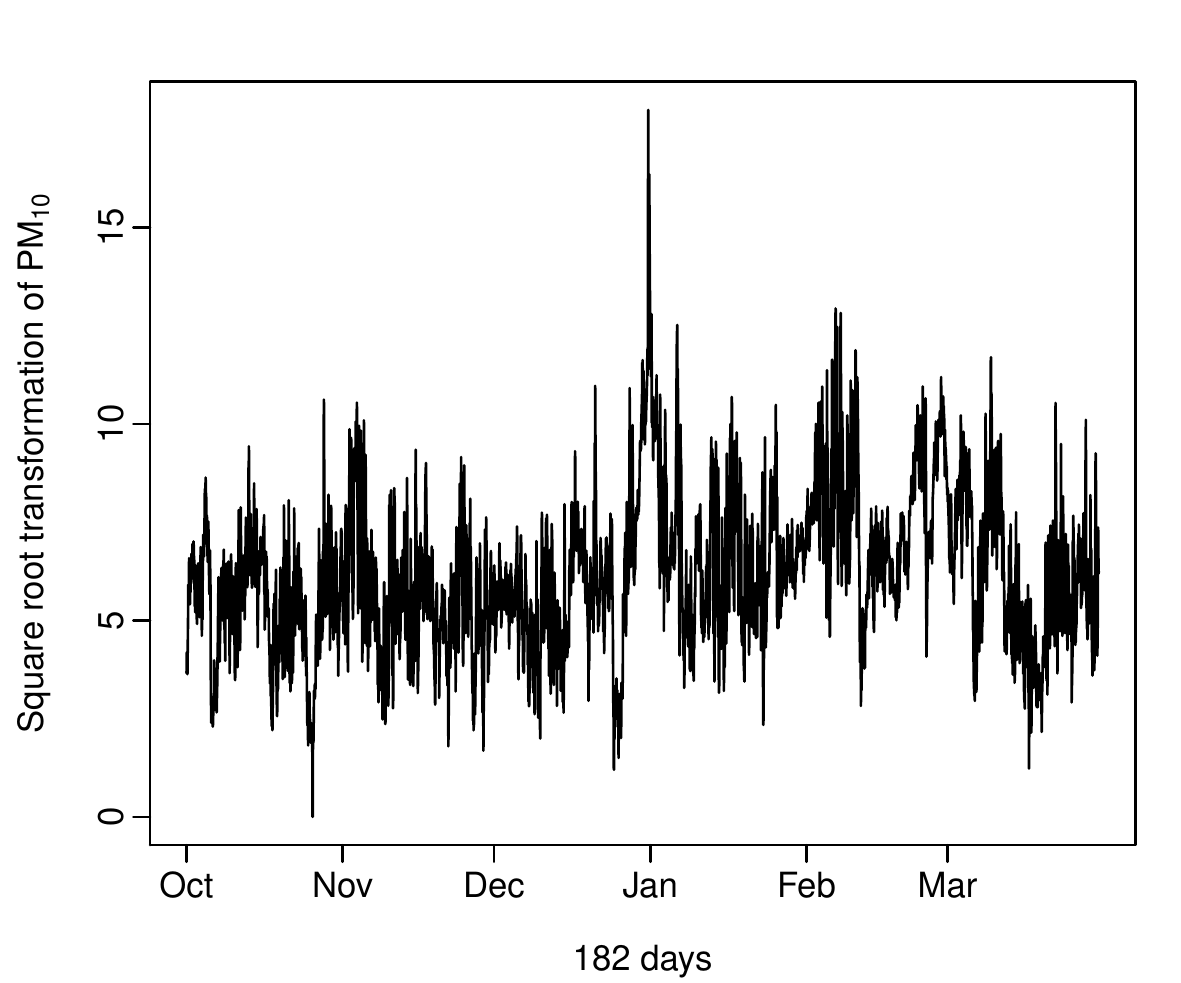}\label{fig:1a}}
\qquad
\subfloat[A functional time series display]
{\includegraphics[width=8.5cm]{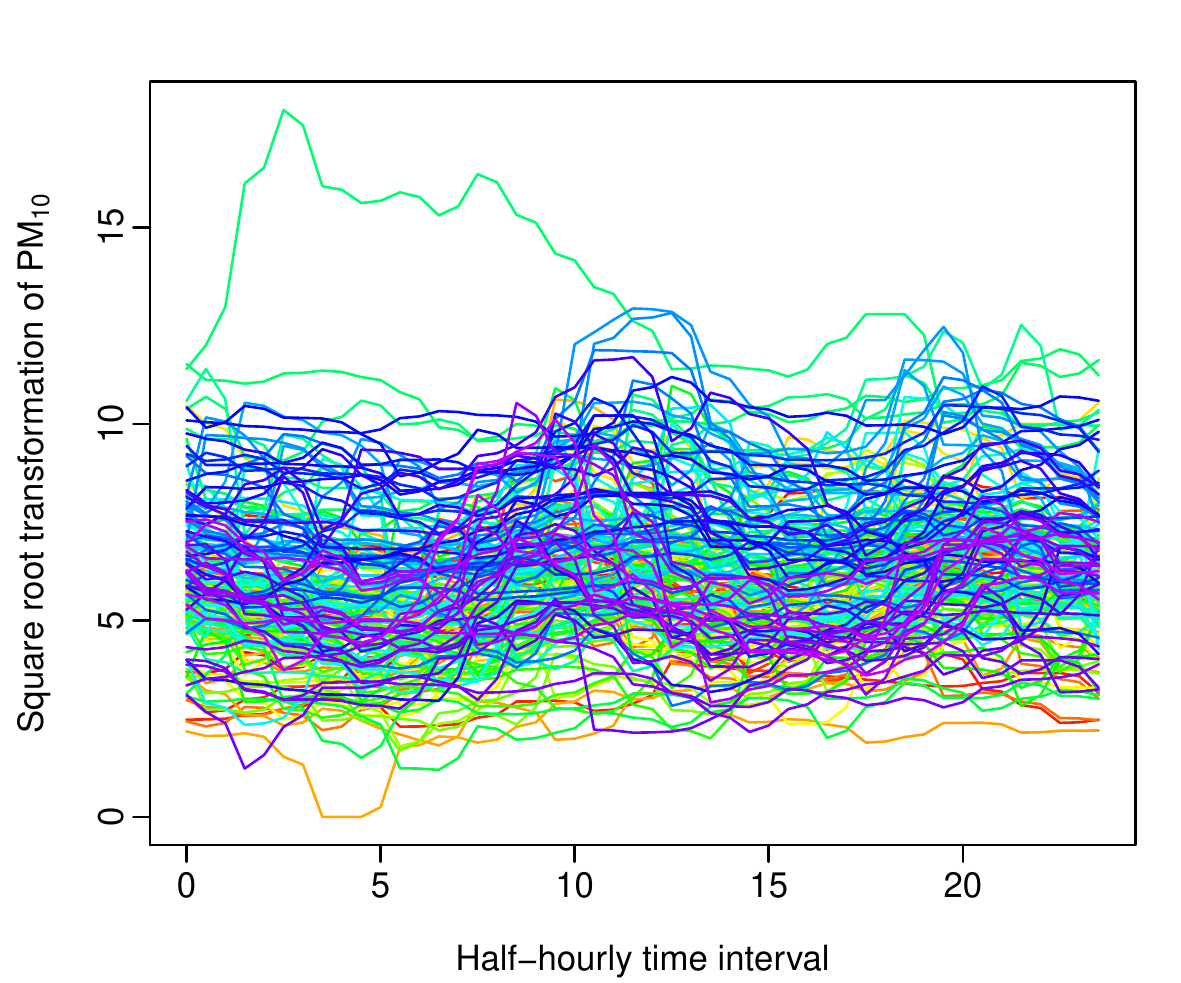}\label{fig:1b}}
\caption{Graphical displays of intraday measurements of the PM$_{10}$ from 1/October/2010 to 31/March/2011.}
\end{figure}

As a vehicle of an illustration, intraday PM$_{10}$ concentrations are considered. The observations are half-hourly measurements of concentration of PM with an aerodynamic diameter of less than $10um$, in ambient air taken in Graz, Austria from 1/October/2010 until 31/March/2011. We convert $N=8,736$ discrete univariate time series points into $n=182$ daily curves. A univariate time series display of intraday pollution curves is given in Figure~\ref{fig:1a}, with the same data shown in Figure~\ref{fig:1b} as a time series of functions.

From Figure~\ref{fig:1b}, there are some days showing extreme measurements of PM$_{10}$. As in a univariate or multivariate time series analysis, the detection of outliers is fundamental in functional time series analysis. According to \cite{FGG07}, a functional outlier is a curve generated by a stochastic process with a different distribution than the one of normal curves. This definition is quite general covering many types of outliers, e.g., magnitude, shape and partial outliers \citep[see][for more details]{SGL15}.

\begin{figure}[!htbp]
\centering
\subfloat[Bivariate HDR boxplot]
{\includegraphics[width=8.5cm]{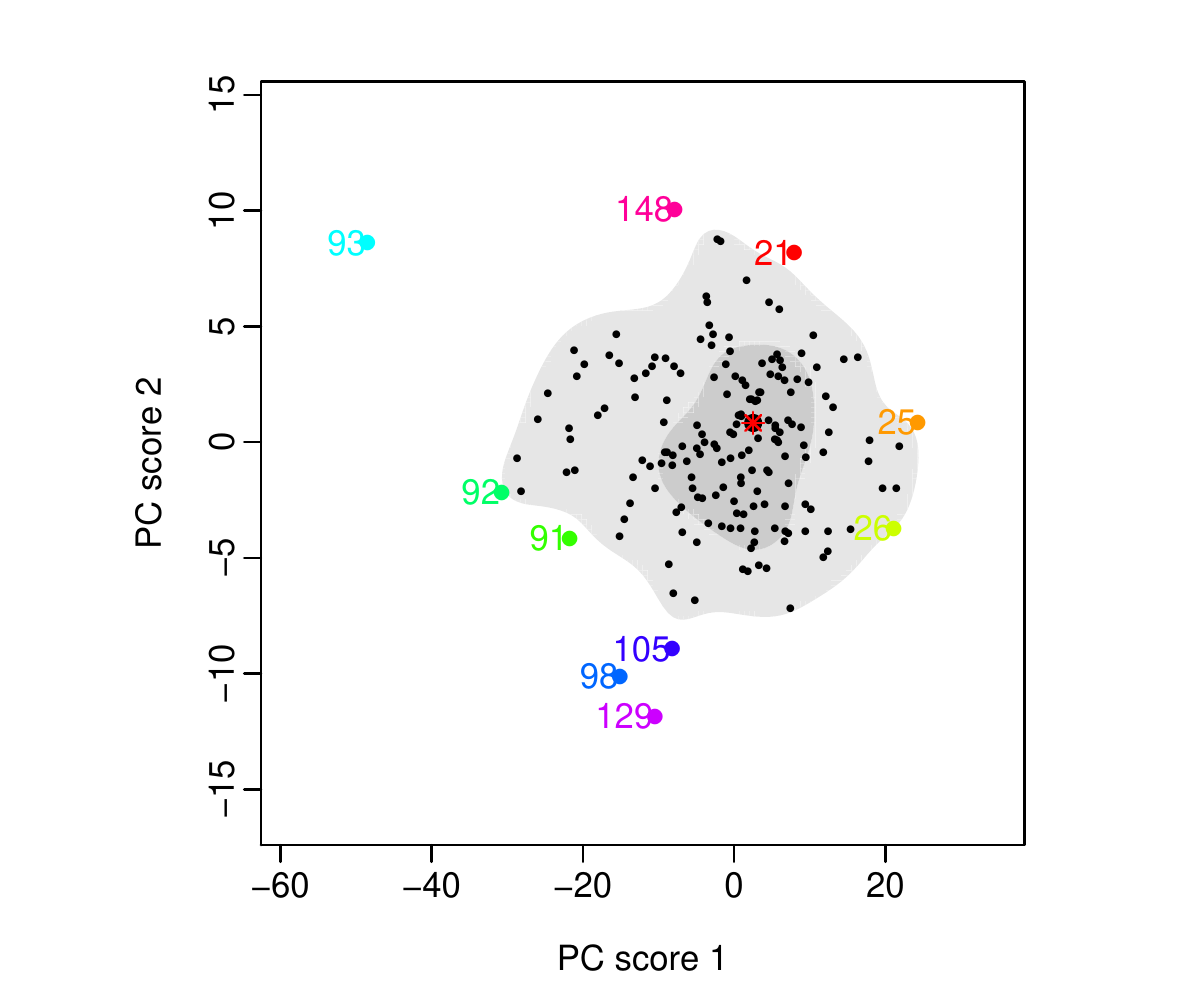}\label{fig:2a}}
\qquad
\subfloat[Functional HDR boxplot (mode is the black line)]
{\includegraphics[width=8.5cm]{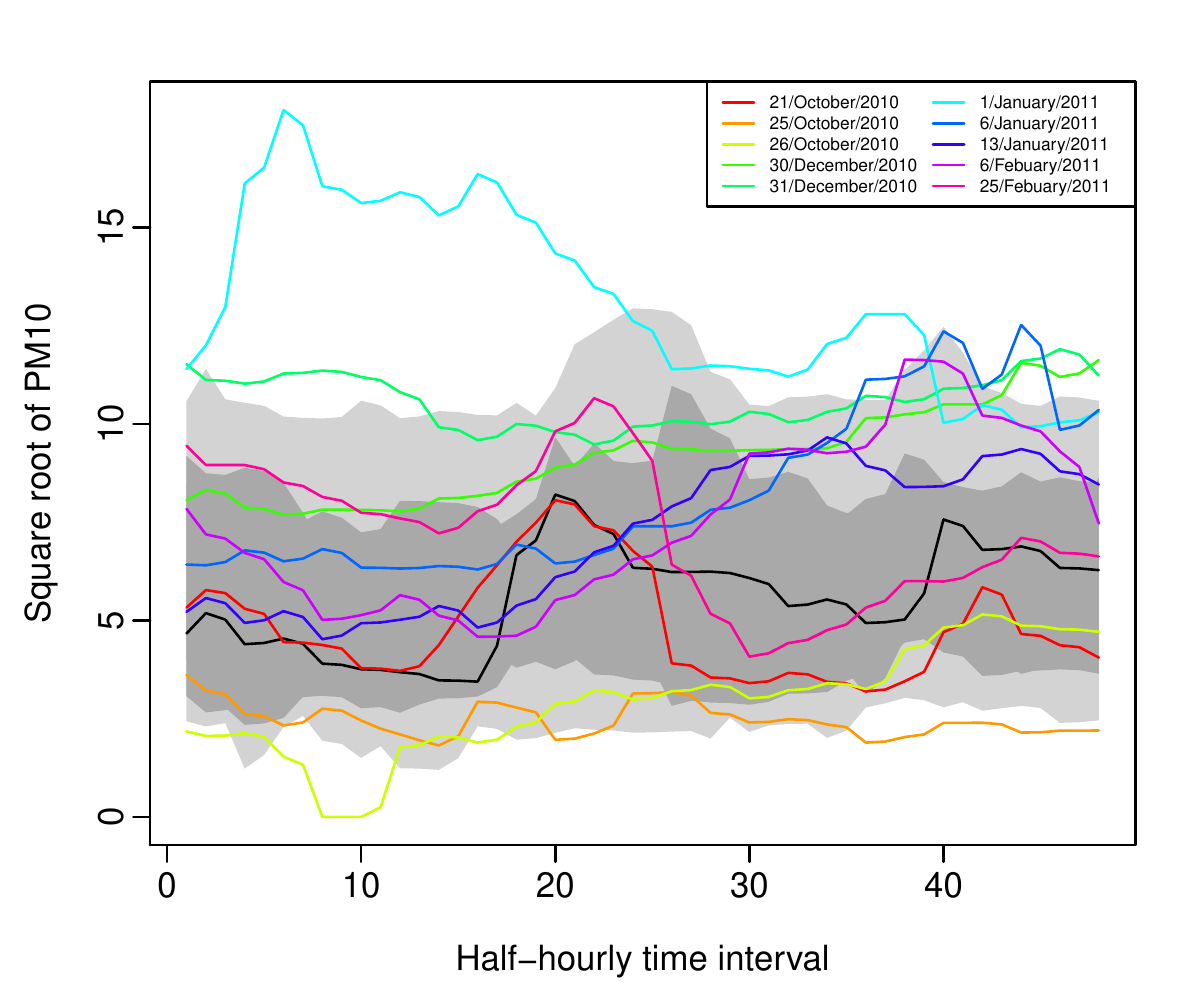}\label{fig:2b}}
\caption{A functional outlier detection method, namely the highest density region, is used to identify ten outliers representing about 5\% of the total number of curves.}\label{fig:2}
\end{figure}

Since the presence of outliers can seriously hinder the performance of modeling and forecasting, we adapted the functional highest density region boxplot of \cite{HS10} and identified ten outliers in Figure~\ref{fig:2}. These outliers correspond to the dates 21/October/2010, 25/October/2010, 26/October/2010, 30/December/2010, 31/December/2010, 1/January/2011, 6/January/2011, 13/January/2011, 6/February/2011 and 25/February/2011, highlighted by the colored lines in Figure~\ref{fig:2b}. In Section~\ref{sec:7}, we compare the forecast accuracy between the standard and robust FPCA, where the latter one is not greatly influenced much by the presence of outliers.

\section{Measures of point and interval forecast accuracy}\label{sec:6}

\subsection{Absolute and squared forecast errors}

We compute the point forecasts between the proposed methods, and evaluate their forecast accuracy by the mean absolute forecast error (MAFE) and mean squared forecast error (MSFE). These both measure how close the forecasts are in comparison to the actual values of the variable being forecast, and these measures can be expressed as
\begin{align*}
\text{MAFE}_j &= \frac{1}{q}\sum^q_{\kappa=1}\left|\mathcal{X}_{n+\kappa}(t_j) - \widehat{\mathcal{X}}_{n+\kappa|n+\kappa-1}(t_j)\right|, \\
\text{MSFE}_j &= \frac{1}{q}\sum^q_{\kappa=1}\left[\mathcal{X}_{n+\kappa}(t_j) - \widehat{\mathcal{X}}_{n+\kappa|n+\kappa-1}(t_j)\right]^2, 
\end{align*}
where $q$ represents the number of curves in the holdout forecasting period, $\mathcal{X}_{n+\kappa}(t_j)$ represents the actual holdout sample for the $j^{\text{th}}$ time period in the $\kappa^{\text{th}}$ curve, while $\widehat{\mathcal{X}}_{n+\kappa}(t_j)$ represents the iterative one-step-ahead point forecasts for the holdout samples. 

\subsection{Interval scores}

In order to evaluate the interval forecast accuracy, we utilize the interval score of \cite{GR07} \citep[see also][]{GK14}. For each year in the forecasting period, the one-step-ahead prediction intervals were calculated at the $(1-\alpha)\times 100\%$ nominal coverage probability. We consider the common case of the symmetric $(1-\alpha)\times 100\%$ pointwise prediction interval, with lower and upper bounds that are predictive quantiles at $\alpha/2$ and $1-\alpha/2$, denoted by $\widehat{\mathcal{X}}^l_{n+\kappa|n+\kappa-1}(t_j)$ and $\widehat{\mathcal{X}}^u_{n+\kappa|n+\kappa-1}(t_j)$. As defined by \cite{GR07}, a scoring rule for the interval forecast at time point $\mathcal{X}_{n+\kappa}(t_j)$ is
\begin{align*}
& S_{\alpha}\left[\widehat{\mathcal{X}}^l_{n+\kappa|n+\kappa-1}(t_j),\widehat{\mathcal{X}}^u_{n+\kappa|n+\kappa-1}(t_j);\mathcal{X}_{n+\kappa}(t_j)\right]  =\left[\widehat{\mathcal{X}}^u_{n+\kappa|n+\kappa-1}(t_j) - \widehat{\mathcal{X}}^l_{n+\kappa|n+\kappa-1}(t_j)\right] + \\
& \hspace{1.5in} \frac{2}{\alpha}\left[\widehat{\mathcal{X}}^l_{n+\kappa|n+\kappa-1}(t_j)-\mathcal{X}_{n+\kappa}(t_j)\right]\mathds{1}\left\{\mathcal{X}_{n+\kappa}(t_j) < \widehat{\mathcal{X}}^l_{n+\kappa|n+\kappa-1}(t_j)\right\} + \\
& \hspace{1.5in} \frac{2}{\alpha}\left[\mathcal{X}_{n+\kappa}(t_j)-\widehat{\mathcal{X}}^u_{n+\kappa|n+\kappa-1}(t_j)\right]\mathds{1}\left\{\mathcal{X}_{n+\kappa}(t_j)>\widehat{\mathcal{X}}_{n+\kappa|n+\kappa-1}^u(t_j)\right\},\quad j=1,\dots,p,
\end{align*}
where $\alpha$ denotes the level of significance, customarily $\alpha=0.2$. The interval score rewards a narrow prediction interval, if and only if the true observation lies within the pointwise prediction interval. The optimal interval score is achieved when $\mathcal{X}_{n+\kappa}(t_j)$ lies between $\widehat{\mathcal{X}}^l_{n+\kappa|n+\kappa-1}(t_j)$ and $\widehat{\mathcal{X}}^u_{n+\kappa|n+\kappa-1}(t_j)$, and the distance between $\widehat{\mathcal{X}}^l_{n+\kappa|n+\kappa-1}(t_j)$ and $\widehat{\mathcal{X}}^u_{n+\kappa|n+\kappa-1}(t_j)$ is minimal. 

Averaged over different days in the forecasting period, the mean interval score for each time point $j$ is defined by
\begin{equation*}
\overline{S}_{\alpha,j} = \frac{1}{q}\sum^{q}_{\kappa=1}S_{\alpha}\left[\widehat{\mathcal{X}}^l_{n+\kappa|n+\kappa-1}(t_j),\widehat{\mathcal{X}}^u_{n+\kappa|n+\kappa-1}(t_j);\mathcal{X}_{n+\kappa}(t_j)\right],
\end{equation*}
where $S_{\alpha}\left[\widehat{\mathcal{X}}^l_{n+\kappa|n+\kappa-1}(t_j),\widehat{\mathcal{X}}^u_{n+\kappa|n+\kappa-1}(t_j);\mathcal{X}_{n+\kappa}(t_j)\right]$ denotes the one-step-ahead interval score at the $\kappa^{\text{th}}$ day of the forecasting period.

\section{Results}\label{sec:7}

\subsection{Simulation study}

In the first simulation study, we compare the finite-sample performance in Section~\ref{sec:7.1} between the standard and robust FPCA, in the presence and absence of contaminating additive outliers. In the second simulation study, we compare the finite-sample performance of the FPCA between the univariate and multivariate time series forecasting methods in Section~\ref{sec:7.2}.

\subsubsection{Comparison between the standard and robust functional principal component analyses}\label{sec:7.1}

We generated an artificial two-dimensional VAR$(2)$ process that obeys the following form:
\begin{equation*}
 \left[ \begin{array}{c}
\beta_1 \\
\beta_2  \end{array} \right]_i = 
\underbrace{\left[ \begin{array}{c}
10 \\
5  \end{array} \right]}_{\bm{B}_0} + 
\underbrace{\left[ \begin{array}{cc}
0.5 & 0.2 \\
-0.2 & -0.5  \end{array} \right]}_{\bm{B}_1}
 \left[ \begin{array}{c}
\beta_1 \\
\beta_2  \end{array} \right]_{i-1}+
\underbrace{\left[ \begin{array}{cc}
-0.3 & -0.7 \\
-0.1 & 0.3  \end{array} \right]}_{\bm{B}_1}
 \left[ \begin{array}{c}
\beta_1 \\
\beta_2  \end{array} \right]_{i-2}+
 \left[ \begin{array}{c}
\varepsilon_1 \\
\varepsilon_2  \end{array} \right]_{i}
\end{equation*}
where $\bm{\varepsilon}_i\sim N_2(\bm{0},\bm{\Sigma})$ denotes an error term, following a bivariate normal distribution with covariance matrix
\begin{equation*}
\bm{\Sigma} = \left(\begin{array}{cc}
1 & 0.2 \\
0.2 & 1  \end{array} \right).
\end{equation*}

To generate additive outliers, we randomly select $m$ bivariate observations, and contaminated them by adding the value 10 to all the components of the selected observations \citep[see also][]{CJ08}. We considered different levels of contamination, ranging from zero to 25 additive outliers for a sample size $n=500$. 

The sample curves were generated using a finite Karhunen-Lo\`{e}ve expansion with the functions $\phi_1(t) = \sin(2\pi t)$, $\phi_2(t) = \cos(2\pi t)$, where $t\in [-1,1]$ denotes a set of equally-spaced 51 grid points. Having simulated contaminated or non-contaminated principal component scores, a set of functional curves was obtained by multiplying with fixed basis functions. That is
\begin{equation*}
\mathcal{X}_i(t) = \beta_{i,1} \phi_1(t) + \beta_{i,2} \phi_2(t),\qquad i=1,\dots,n.
\end{equation*}

\begin{figure}[!htbp]
\centering
\subfloat[Simulated non-contaminated curves]
{\includegraphics[width=8.5cm]{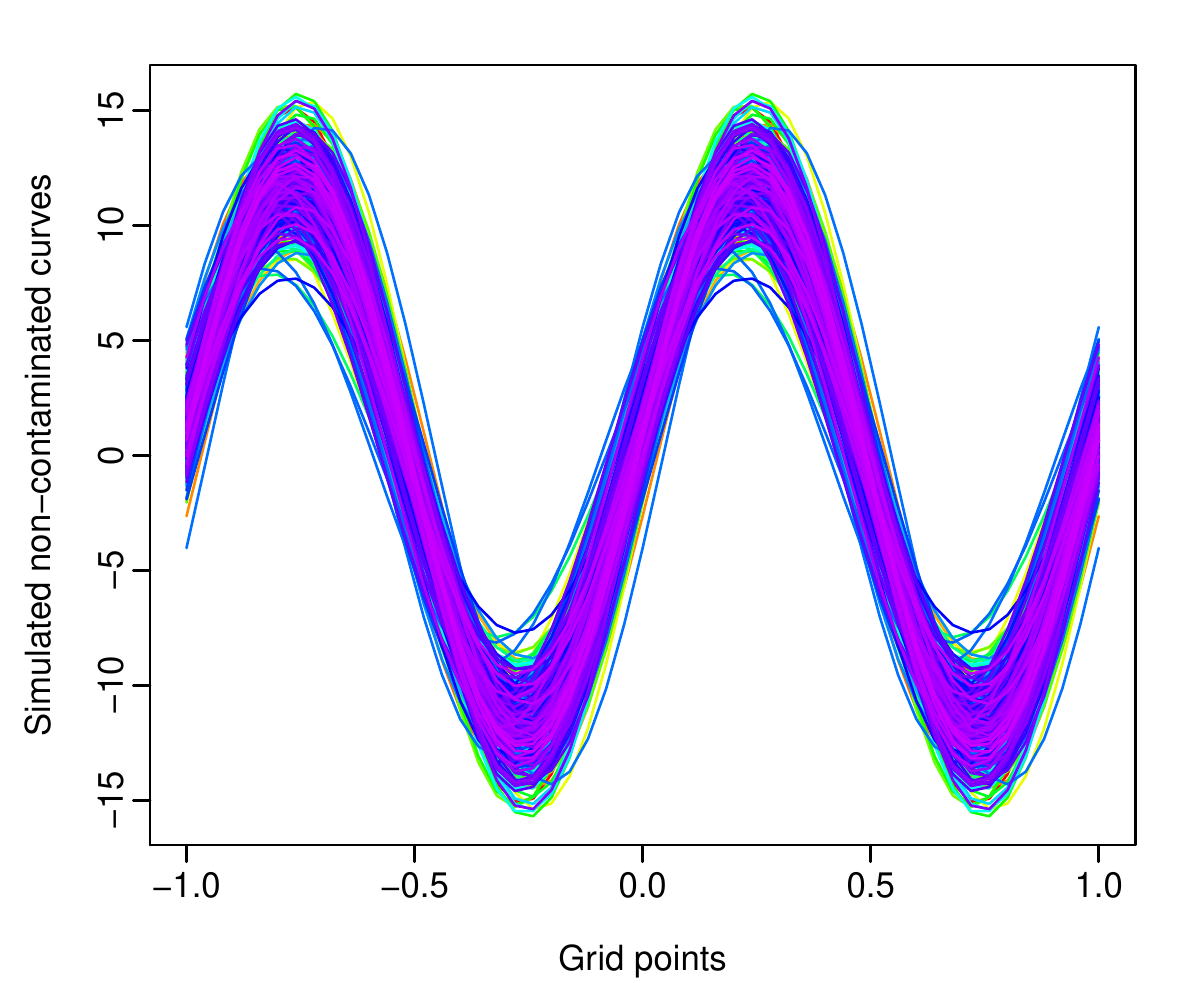}}
\qquad
\subfloat[Simulated contaminated curves ($m=25$)]
{\includegraphics[width=8.5cm]{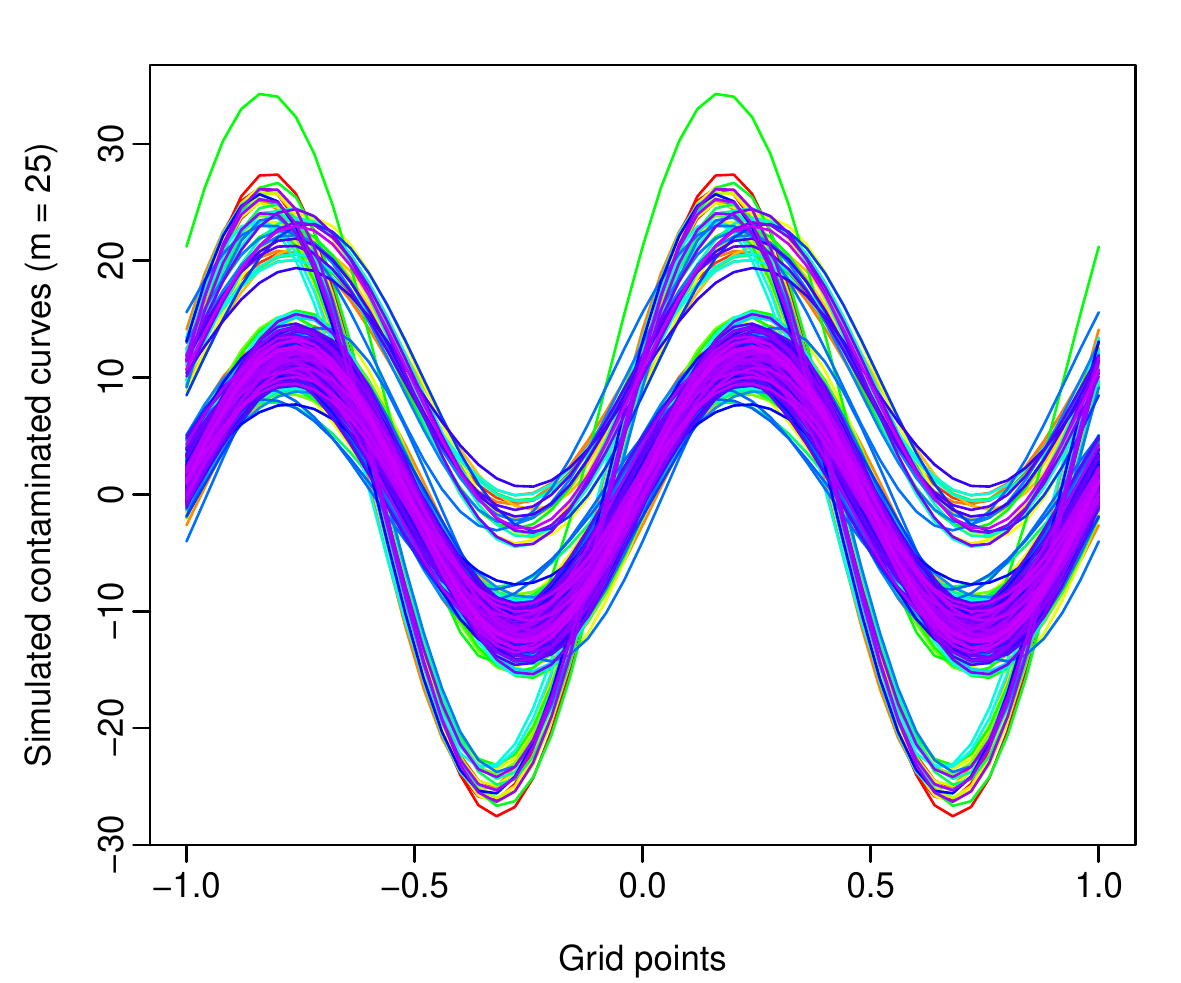}}
\caption{Simulated functional curves. Outliers can stem from both simulated principal component scores and functional curves.}\label{fig:sim_curve}
\end{figure}

Outliers can stem from simulated principal component scores, but can also stem from functional outliers. To generate additive functional outliers, we randomly select $m$ curves, and contaminated them by adding the value 10 to the selected curves. We also consider different levels of contamination, ranging from zero to 25 additive functional outliers. In Figure~\ref{fig:sim_curve}, we present 500 non-contaminated functional curves, as well as contaminated functional curves with $m=25$ outliers.

The goal is to examine the effect of the outliers on the one-step-ahead point forecast accuracy. We first simulate 501 functional curves without contamination, and keep the last curve as a testing sample. Using the first 500 simulated curves, we estimate parameters in the optimal VAR model, produce one-step-ahead point forecast and compare its point forecast accuracy with the testing sample. In order to introduce outliers, the first 500 simulated curves were also simulated with different levels of contamination.

The MAFE and MSFE are used to assess forecast accuracy, and they can be defined as
\begin{equation*}
\text{MAFE} = \frac{1}{51}\sum_{j=1}^{51}\left|\mathcal{X}_{n+1}(t_j) - \widehat{\mathcal{X}}_{n+1|n}(t_j)\right|, \qquad
\text{MSFE} = \frac{1}{51}\sum_{j=1}^{51}\left[\mathcal{X}_{n+1}(t_j) - \widehat{\mathcal{X}}_{n+1|n}(t_j)\right]^2.
\end{equation*}
For $1000$ replications, we use the median of MSFE and MAFE as overall forecast error measures. The one-step-ahead median MSFE and MAFE for the standard and robust FPCA are given in Table~\ref{tab:simu_1}, as a function of the number of outliers out of 500 simulated curves. 

\begin{table}[!htbp]
\tabcolsep 0.39in
\centering
\begin{tabular}{@{}lcccc@{}}\toprule
\# of outliers & \multicolumn{2}{c}{MAFE} & \multicolumn{2}{c}{MSFE} \\
& FPCA & Robust FPCA & FPCA & Robust FPCA \\\midrule
0   & \textBF{0.7681} & 0.7684 & 0.7259 & \textBF{0.7256} \\
1   & \textBF{0.7914} & 0.7919 & 0.7769 & \textBF{0.7738} \\
2   & \textBF{0.8165} & 0.8190 & \textBF{0.8217} & 0.8249 \\
3   & 0.8406               & \textBF{0.8374} & 0.8736 & \textBF{0.8634} \\
4   & 0.8554               & \textBF{0.8515} & 0.9111 & \textBF{0.8930} \\
5   & 0.8626               & \textBF{0.8557} & 0.9244 & \textBF{0.9053} \\
10 & 1.0846               & \textBF{0.9073} & 1.4733 & \textBF{1.0131} \\
15 & 1.1049               & \textBF{0.9341} & 1.5554 & \textBF{1.0761} \\
20 & 1.1267               & \textBF{0.9865} & 1.6383 & \textBF{1.2032} \\
25 & 1.1838               & \textBF{1.0360} & 1.8636 & \textBF{1.3241} \\\bottomrule
\end{tabular}
\caption{Point forecast accuracy comparison between the standard and robust FPCA.} \label{tab:simu_1}
\end{table}

We found the following evidences:
\begin{asparaenum}
\item[(a)] In the presence of outliers, the standard and robust FPCA perform similarly.
\item[(b)] When the number of outliers equals $\{3, 4\}$, the robust FPCA performs slightly better than the standard FPCA.
\item[(c)] When the number of outliers equals $\{5, 10, 15, 20, 25\}$, the robust FPCA performs better than the standard FPCA.
\end{asparaenum}
From the above findings, the robust FPCA outperforms the standard FPCA as the number of outliers increases.

\subsubsection{Comparison between the univariate and multivariate time series forecasting methods}\label{sec:7.2}

We consider the case of non-contamination, where $\bm{\beta} = [\bm{\beta}_1, \bm{\beta}_2]^{\top}$ were generated via the VAR(2). In Table~\ref{tab:simu_2}, we report the one-step-ahead median MAFE and MSFE in 1000 replications between the ARIMA and VAR forecasting methods, and found that the VAR outperforms the ARIMA. These results, similar to those obtained by \cite{SNR13}, suggest that multivariate Generalized Autoregressive Conditional Heteroskedasticity models outperformed competing univariate models on an out-of-sample basis.

\begin{table}[!htbp]
\centering
\begin{tabular}{@{}lll@{}}\toprule
Error & ARIMA & VAR \\\midrule
MAFE & 1.0640 & \textBF{0.7681} \\
MSFE & 1.3952 & \textBF{0.7259}  \\\bottomrule
\end{tabular}
\caption{Point forecast accuracy comparison between the ARIMA and VAR forecasting methods.}\label{tab:simu_2}
\end{table}

\subsection{PM$_{10}$ data set}

\begin{figure}[!htbp]
\centering
\includegraphics[width=6.2cm]{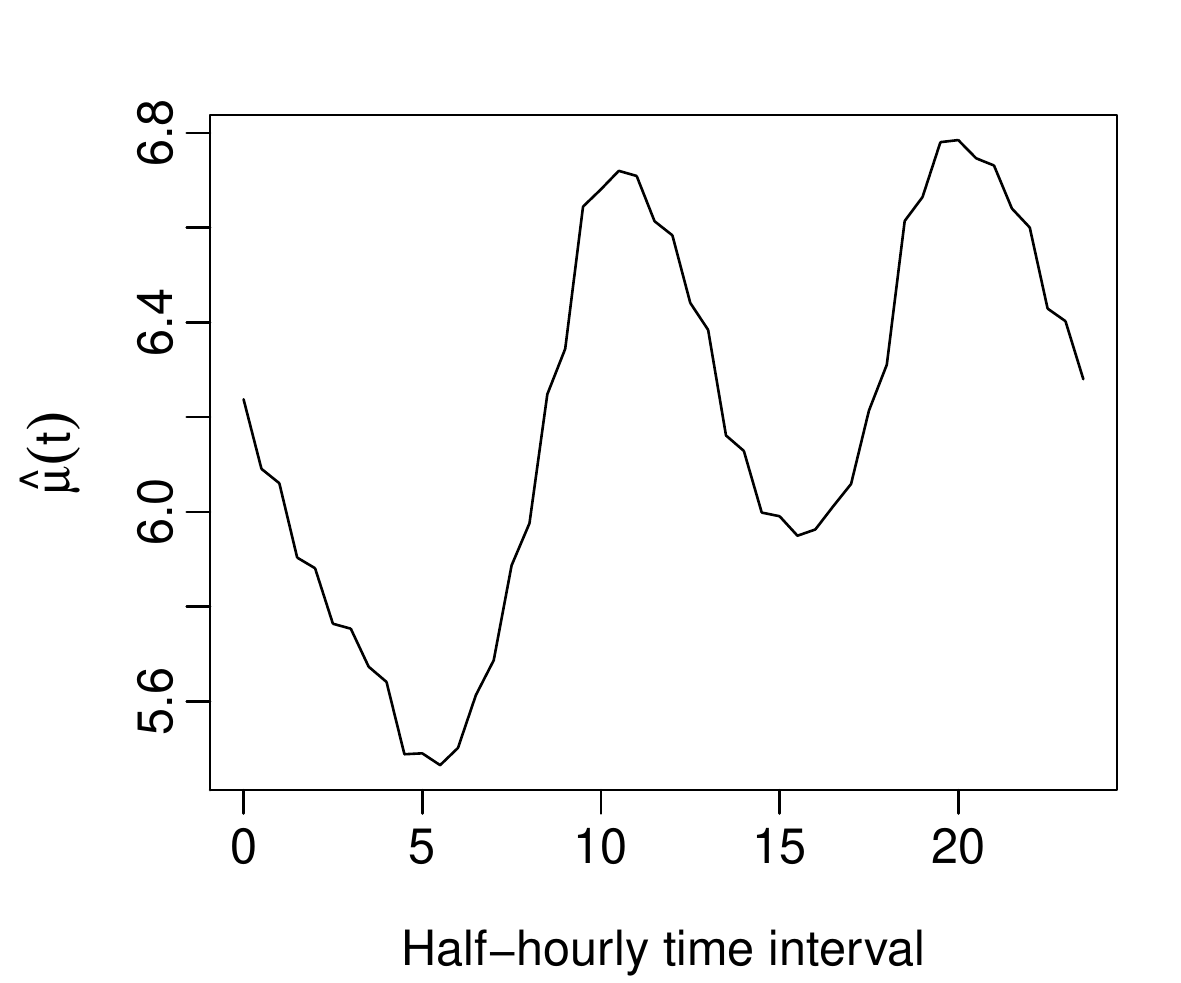}
\quad
\includegraphics[width=6.2cm]{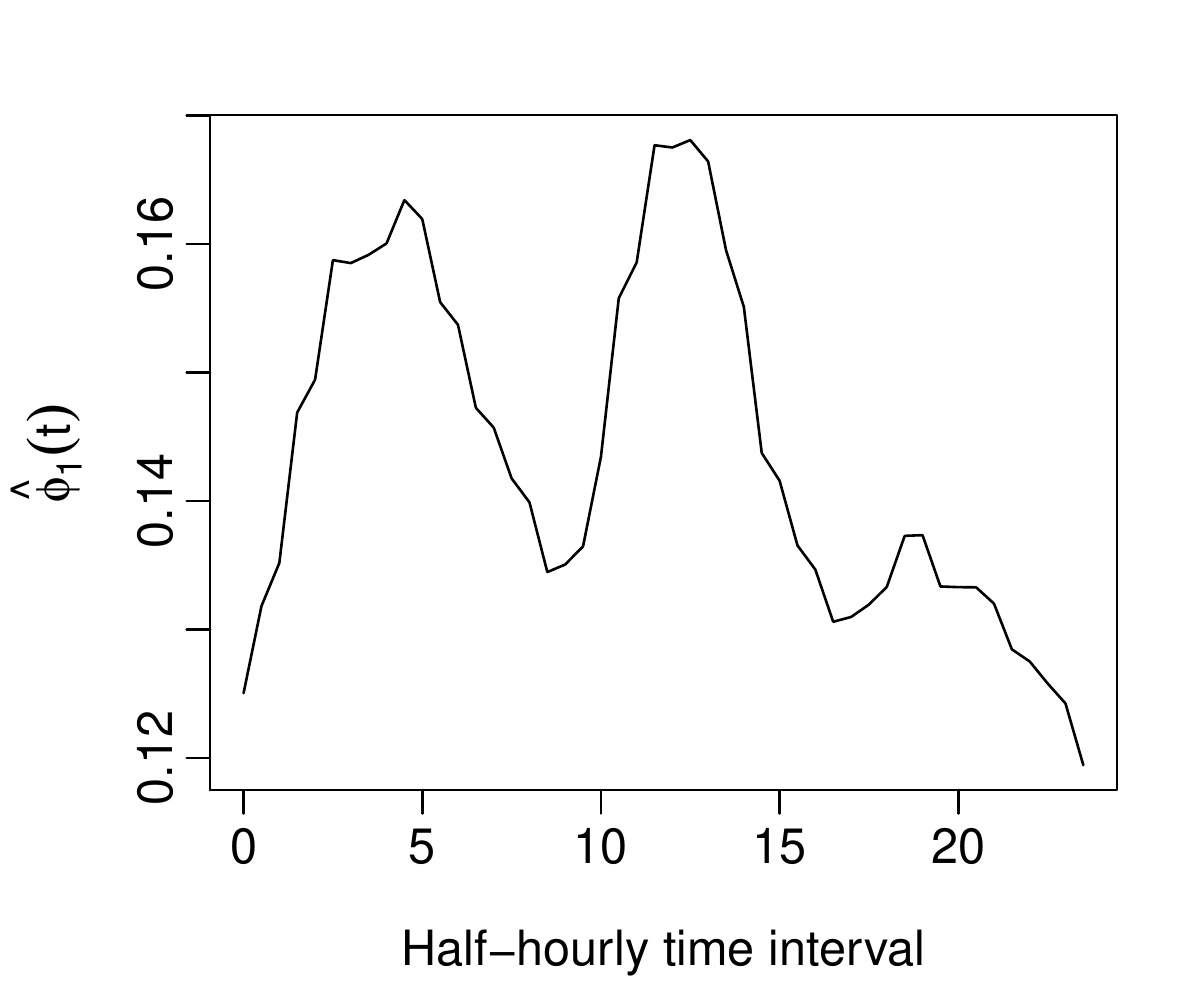}
\\
\includegraphics[width=6.2cm]{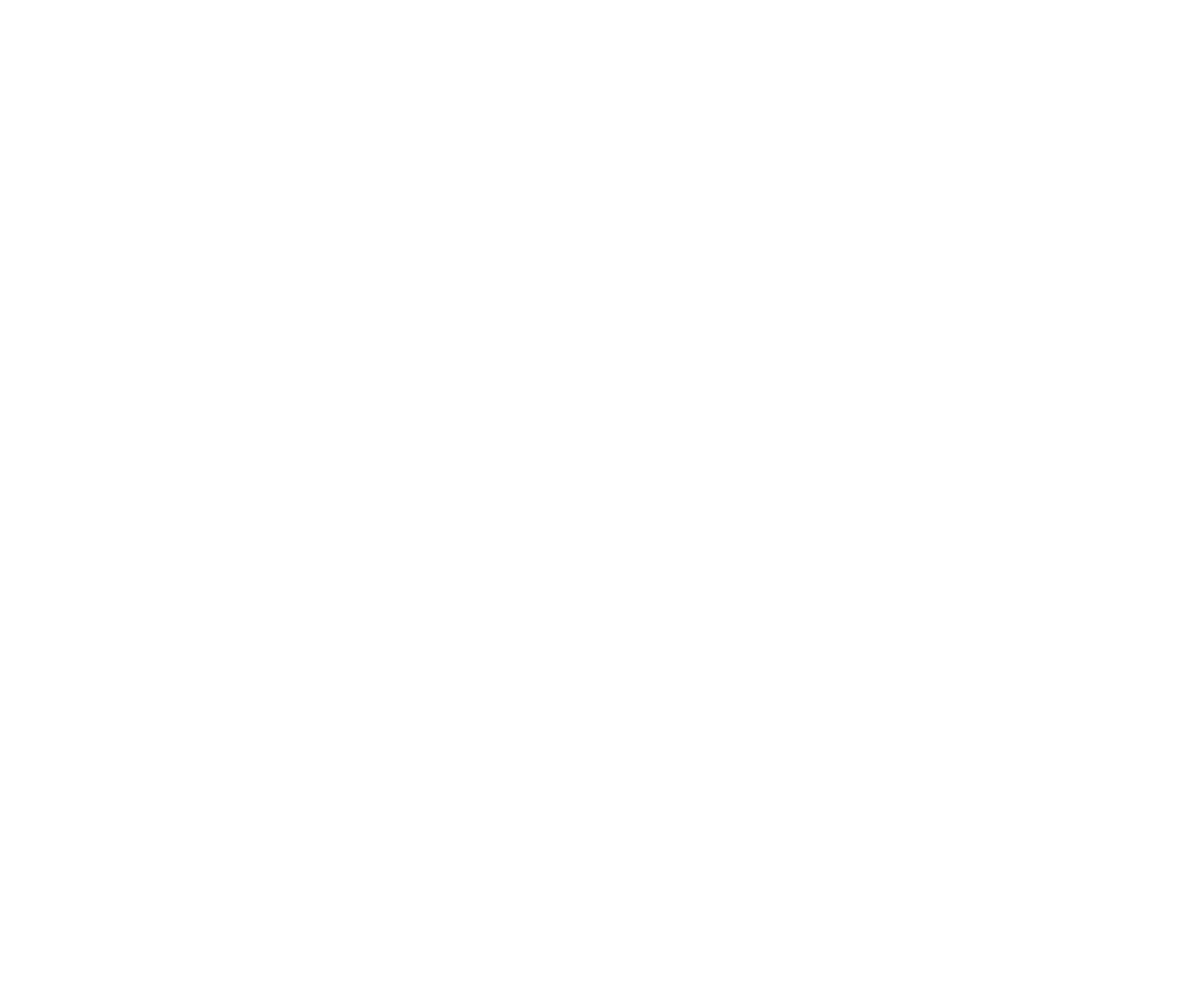}
\quad
\includegraphics[width=6.2cm]{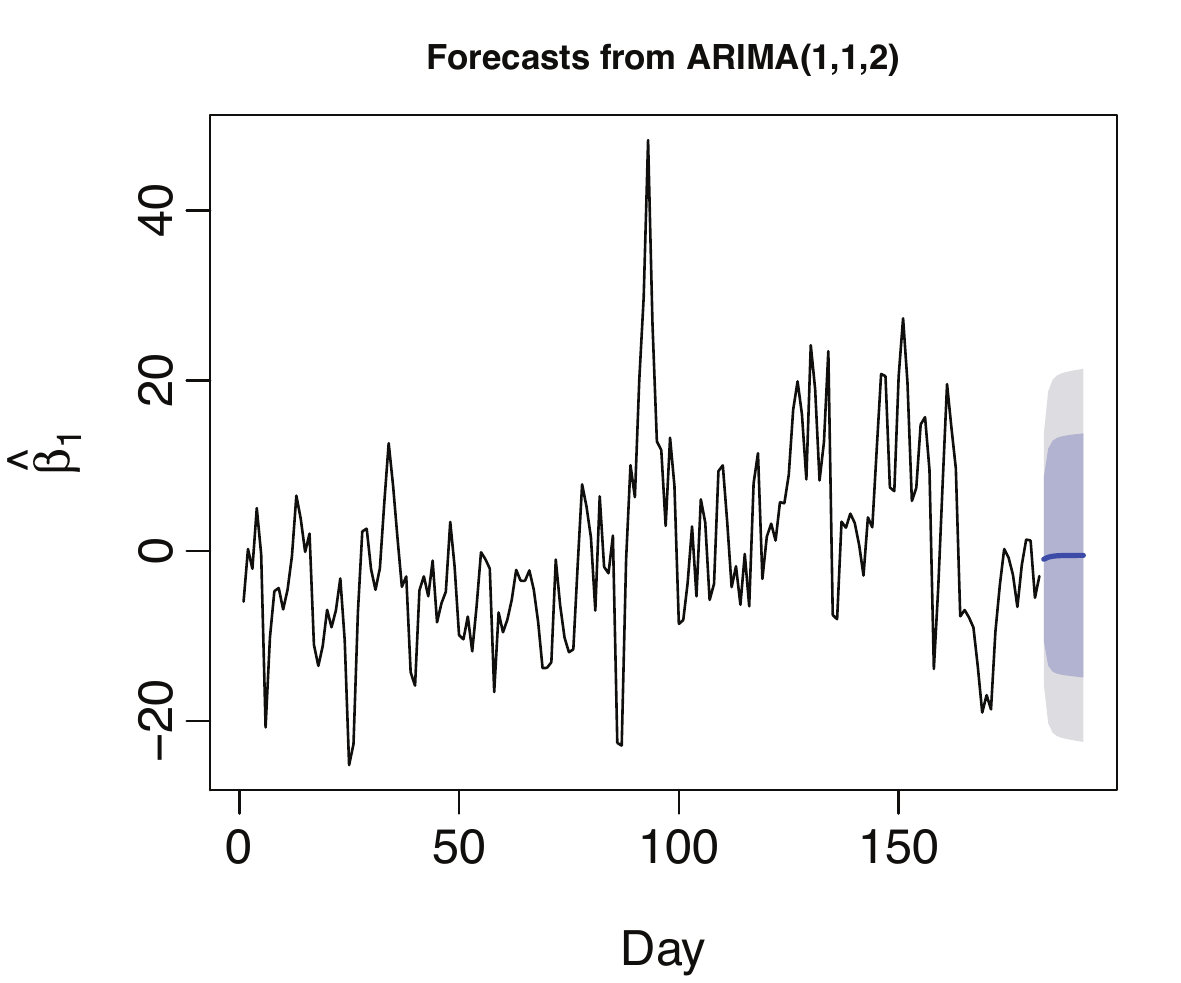}
\caption{The estimated mean function, first functional principal component and its associated scores for the intraday PM$_{10}$ measurements from 1/October/2010 to 31/March/2011. For ten-days-ahead forecasts, the 80\% and 95\% prediction intervals of the estimated principal component scores using the ARIMA method are shown by the dark and light gray regions.}\label{fig:pca_decomp}
\end{figure}

Our forecasting method decomposes a functional time series into a number of functional principal components and their associated scores. The temporal dependency of the functional time series is inherited in the temporal dependency of the principal component scores. For ease of presentation, we display and attempt to interpret only the first functional principal component in the top panel of Figure~\ref{fig:pca_decomp}, although the number of retained principal components is determined by explaining at least 90\% of the total variation in the data. The mean function illustrates the average changes in intraday PM$_{10}$, with two peaks occurring at 11am and 8pm, and two troughs occurring at 5:30am and 3:30pm. The first functional principal component shows contrasts between early and late morning, as well as early and late afternoon. Using a univariate time series forecasting technique, the forecast of principal component scores for ten-days-ahead show a slight increasing trend before stabilizing. This reflects the short-term prediction ability of intraday PM$_{10}$ curves.

\begin{figure}[!htbp]
\centering
\includegraphics[width=9cm]{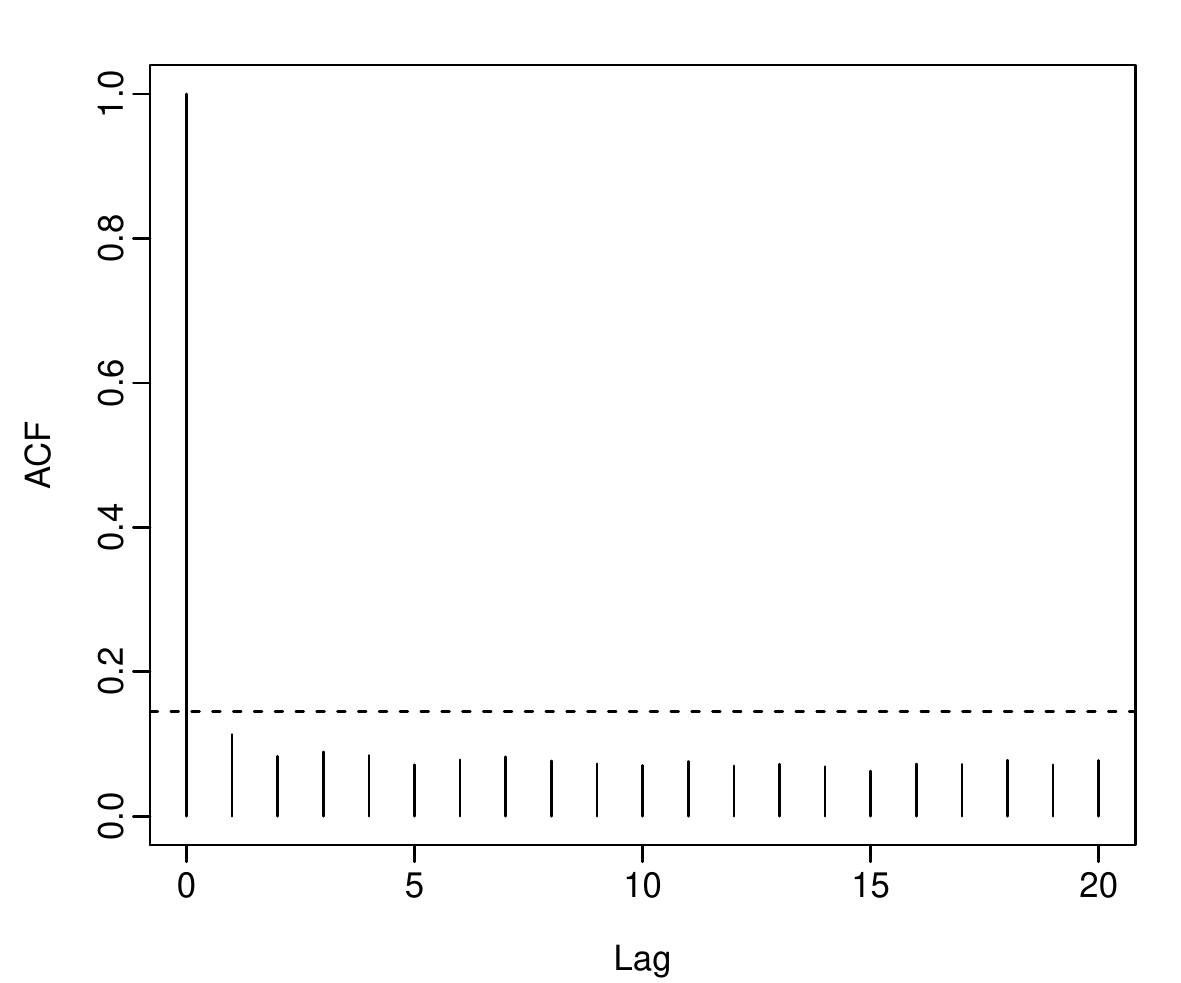}
\caption{Functional ACF plot. The blue dotted line symbolizes the critical value ($1.96/\sqrt{n}$).}\label{fig:facf}
\end{figure}

To assess the overall goodness of fit, we focus on the residual functions of the fitted functional time series model using the ARIMA forecasting method. Following the early work by \cite{HRW16}, we compute the functional analogue of autocorrelation function (ACF) to examine if there is any remaining temporal dependency in Figure~\ref{fig:facf}. The functional ACF is defined as
\begin{equation*}
\widehat{\rho}_i = \frac{\|\widehat{\gamma}_i\|}{\int \widehat{\gamma}_0(t,t)dt},
\end{equation*}
where $\|\cdot\|$ denotes the $L_2$ norm. Since $\|\widehat{\gamma}_i\|\geq 0$, $\widehat{\rho}_i>0$ measures only the strength of the temporal dependency. Since the functional ACFs at all lags that are greater than zero are less than the critical value ($1.96/\sqrt{n}$), we conclude that there is no temporal dependency in these residual functions. Furthermore, using the stationarity test of \cite{HKR14}, we found that the time series of the residual functions are stationary, with a $p$-value of 0.111. 

\begin{figure}[!htbp]
\centering
\includegraphics[width=8.5cm]{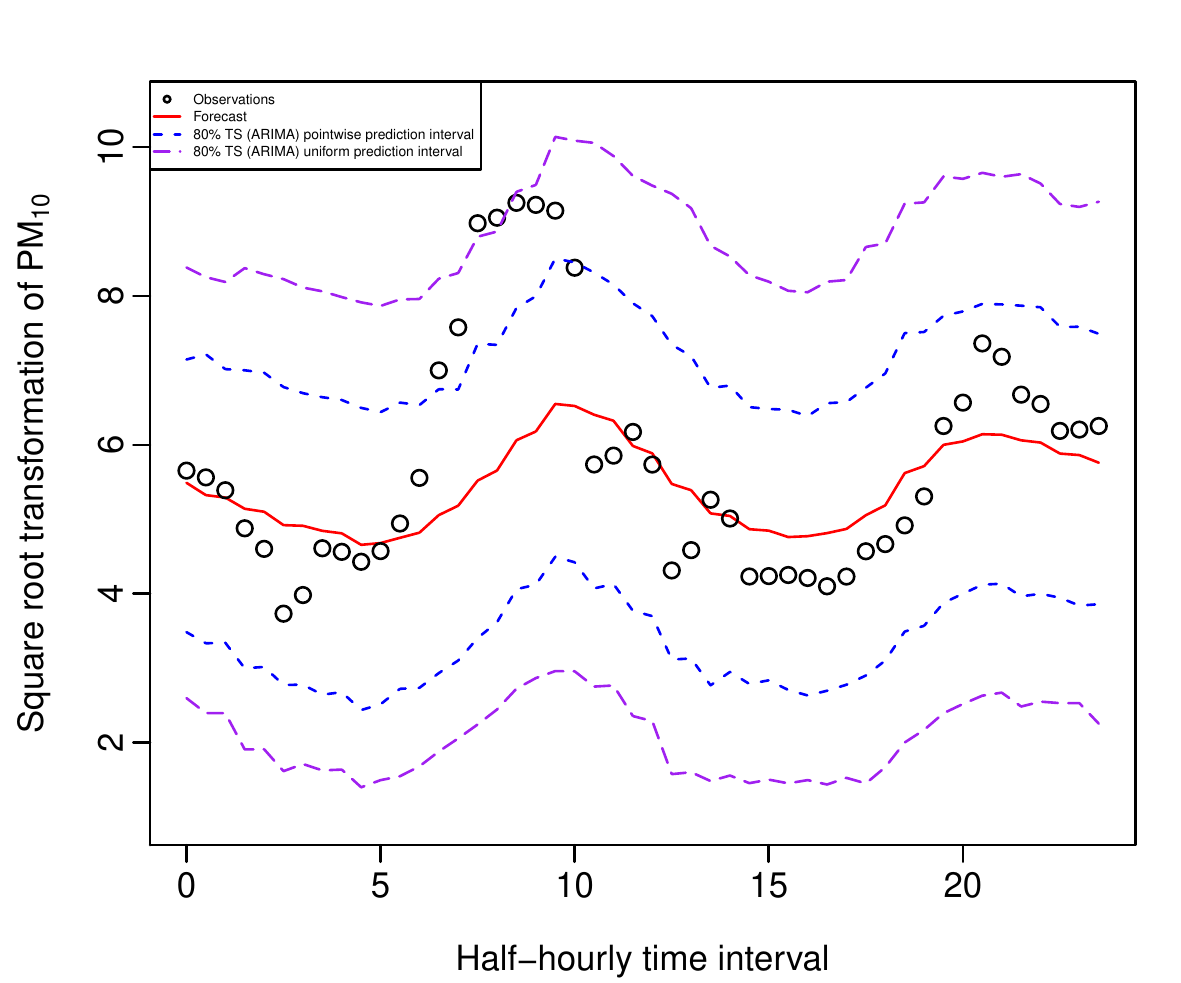}
\quad
\includegraphics[width=8.5cm]{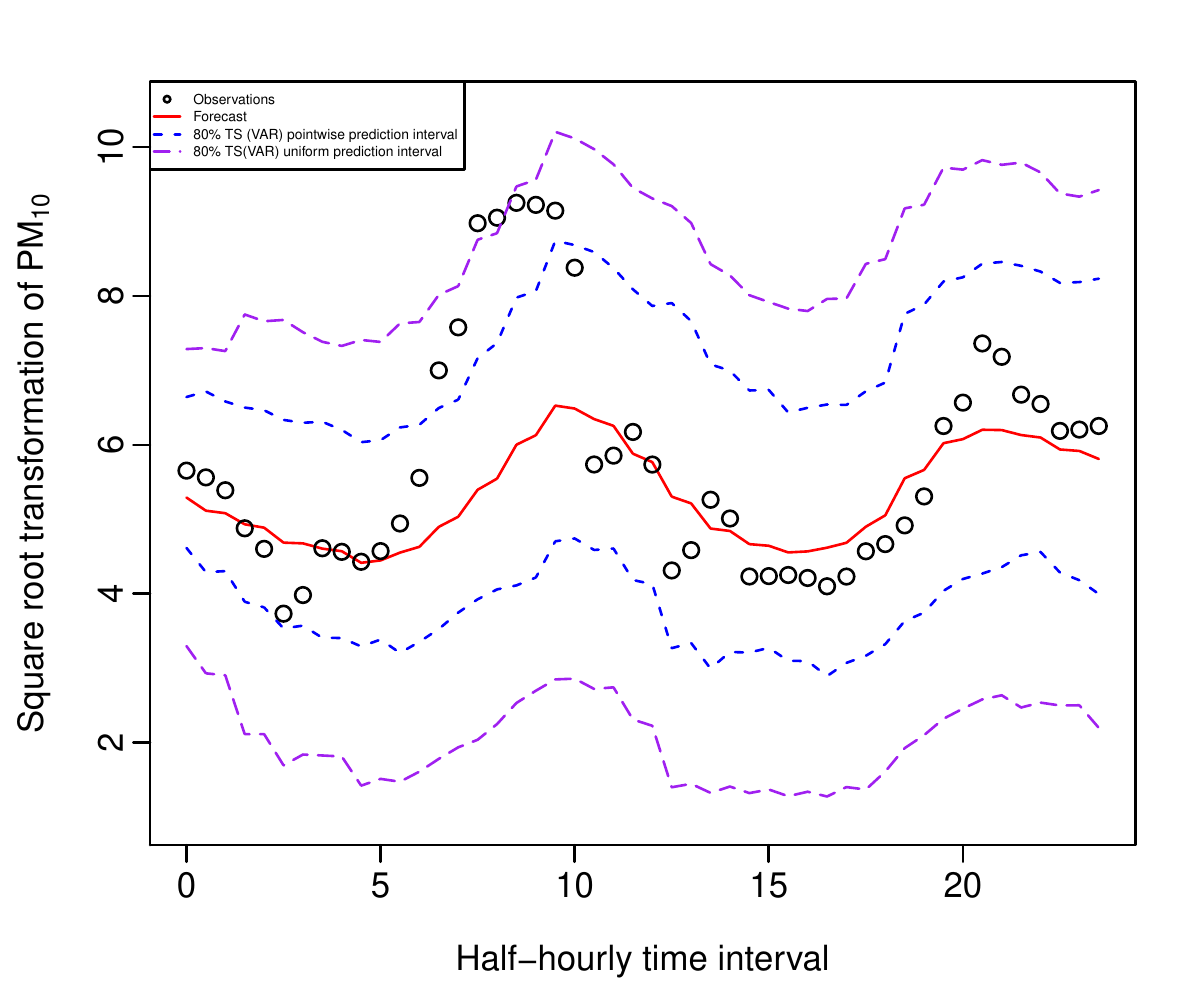}
\caption{Point forecasts of the intraday PM$_{10}$ on 31/March/2011, and the 80\% pointwise and uniform prediction intervals constructed via the nonparametric bootstrap method and prediction band.}\label{fig:point_fore}
\end{figure}

As an illustration, the one-day-ahead point forecast of intraday PM$_{10}$ curve is obtained by multiplying the forecast principal component scores by the estimated functional principal components, and adding the estimated mean function. Through the nonparametric bootstrapping and prediction band described in Sections~\ref{sec:5.1} and~\ref{sec:band} respectively, the 80\% pointwise and uniform prediction intervals are constructed and presented in Figure~\ref{fig:point_fore}.

\begin{figure}[!htbp]
\centering
\includegraphics[width=5.9cm]{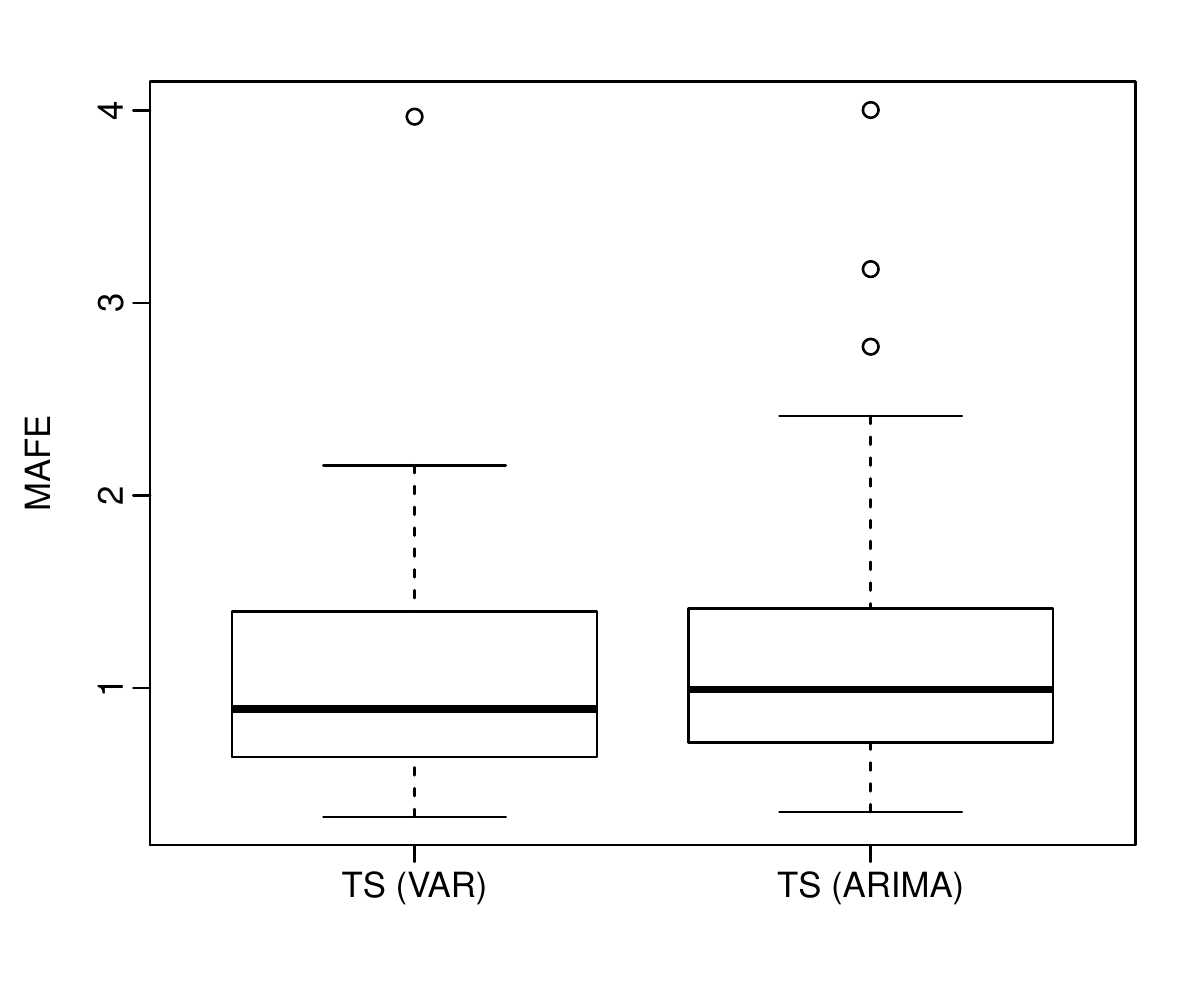}
\includegraphics[width=5.9cm]{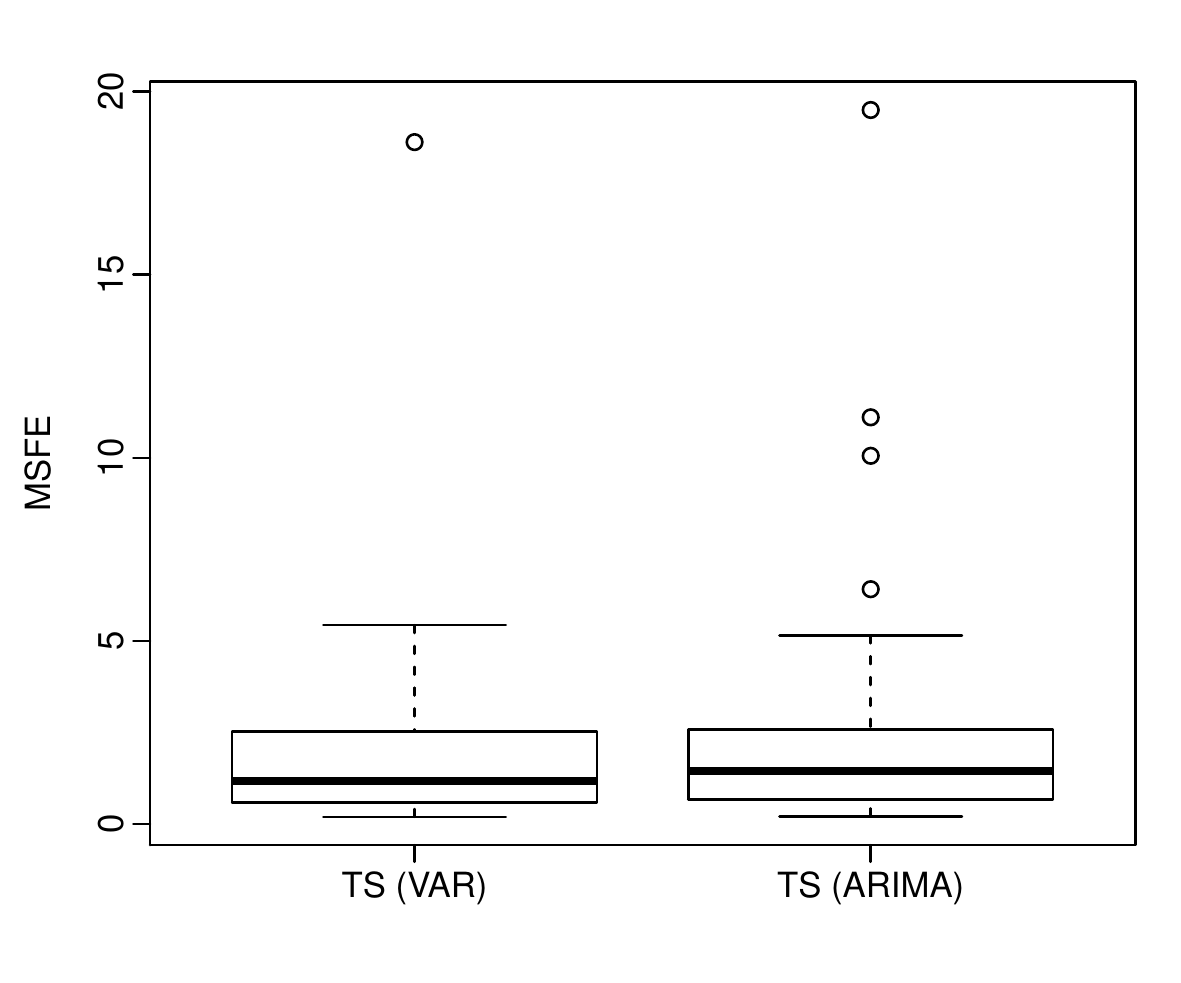}
\includegraphics[width=5.9cm]{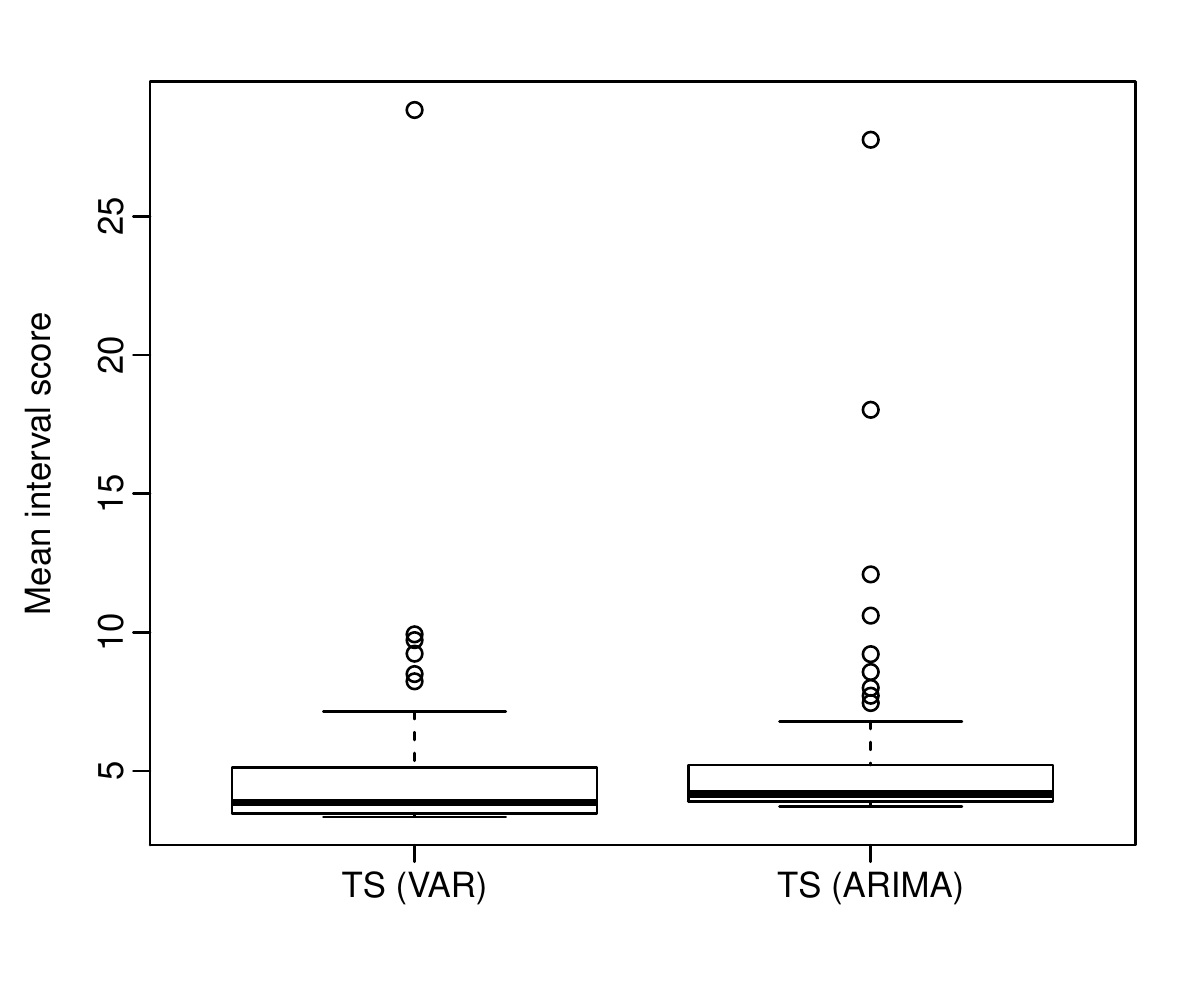}
\caption{Comparison of the point and interval forecast accuracies generated by the ARIMA and VAR forecasting methods over the intraday PM$_{10}$ curves for the forecasting period between the ARIMA and VAR forecasting methods.}\label{fig:uni_var}
\end{figure}

We investigate the point and interval forecast accuracies of the functional principal component regression with univariate and multivariate time series forecasting techniques in Figure~\ref{fig:uni_var}. For this data set, we find that the VAR forecasting method produces smaller forecast errors than the ARIMA forecasting method. 

Averaged over the last 72 days in the forecasting period, the averaged MSFE is 2.24, the averaged MAFE is 1.14, and averaged mean interval score is 5.42, for the TS method with the ARIMA forecasting technique. For the TS method with the VAR forecasting technique, the averaged MSFE is 1.92, the averaged MAFE is 1.06, and the averaged mean interval score is 4.92. 

\begin{figure}[!htbp]
\centering
\includegraphics[width=5.9cm]{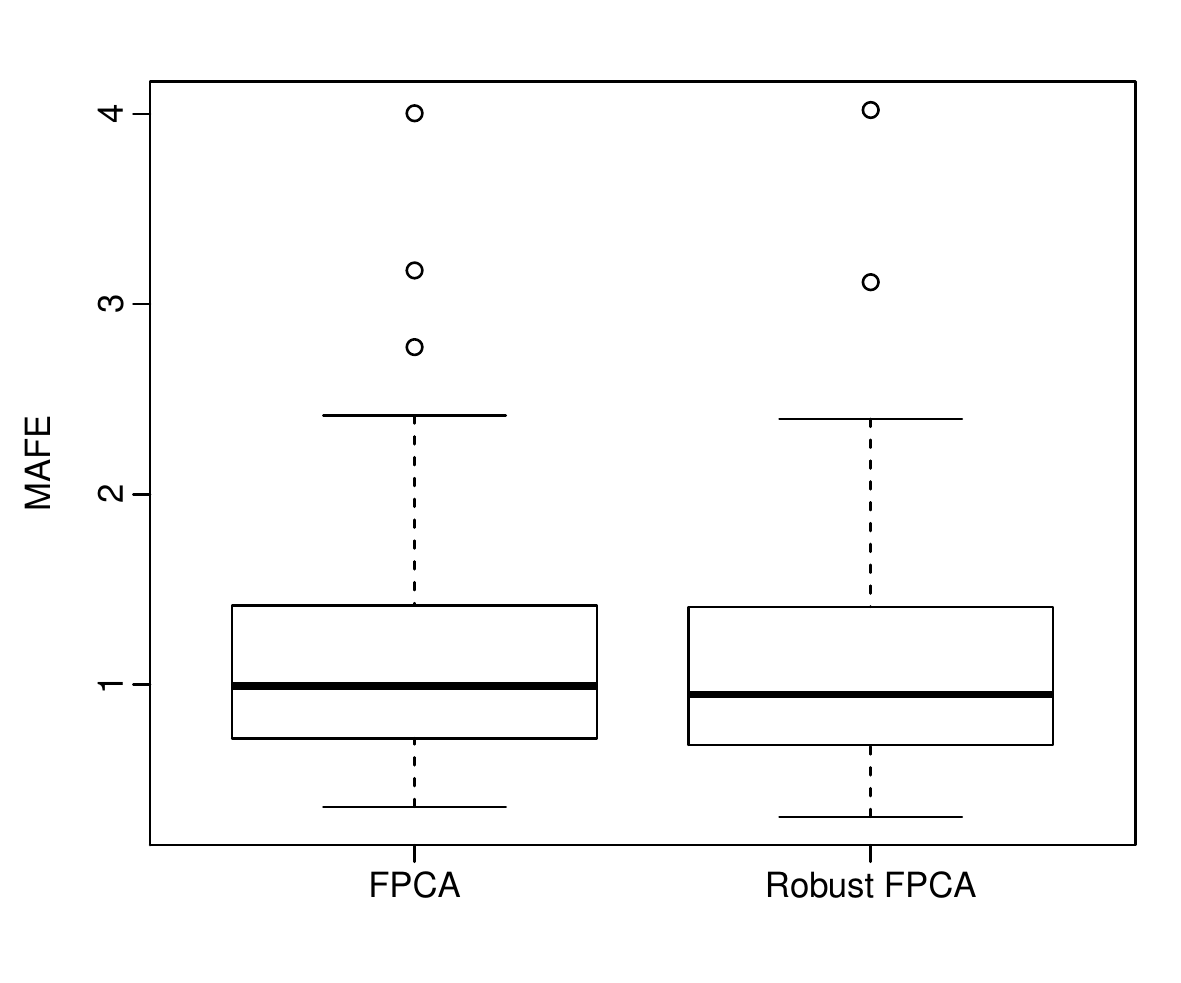}
\includegraphics[width=5.9cm]{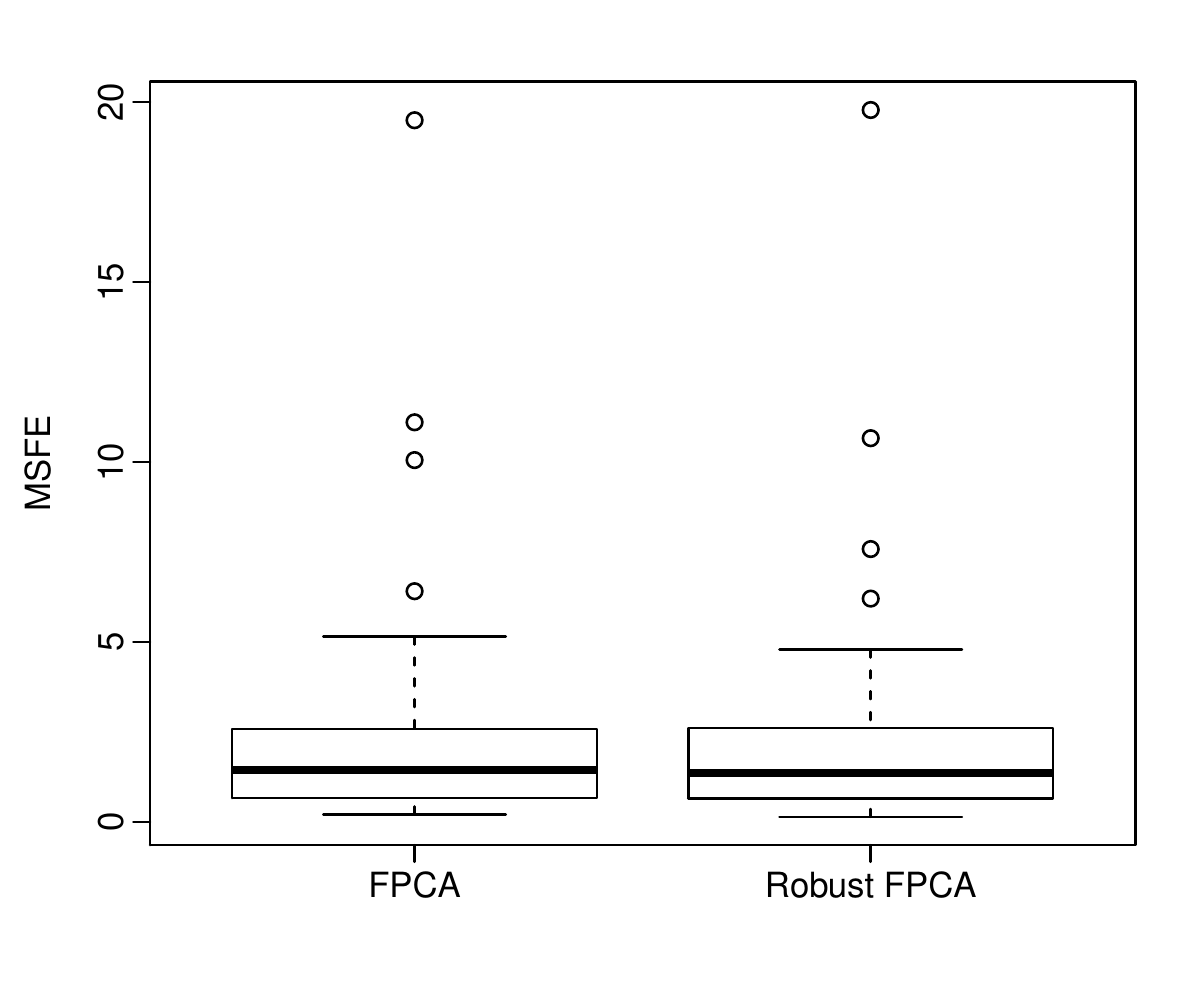}
\includegraphics[width=5.9cm]{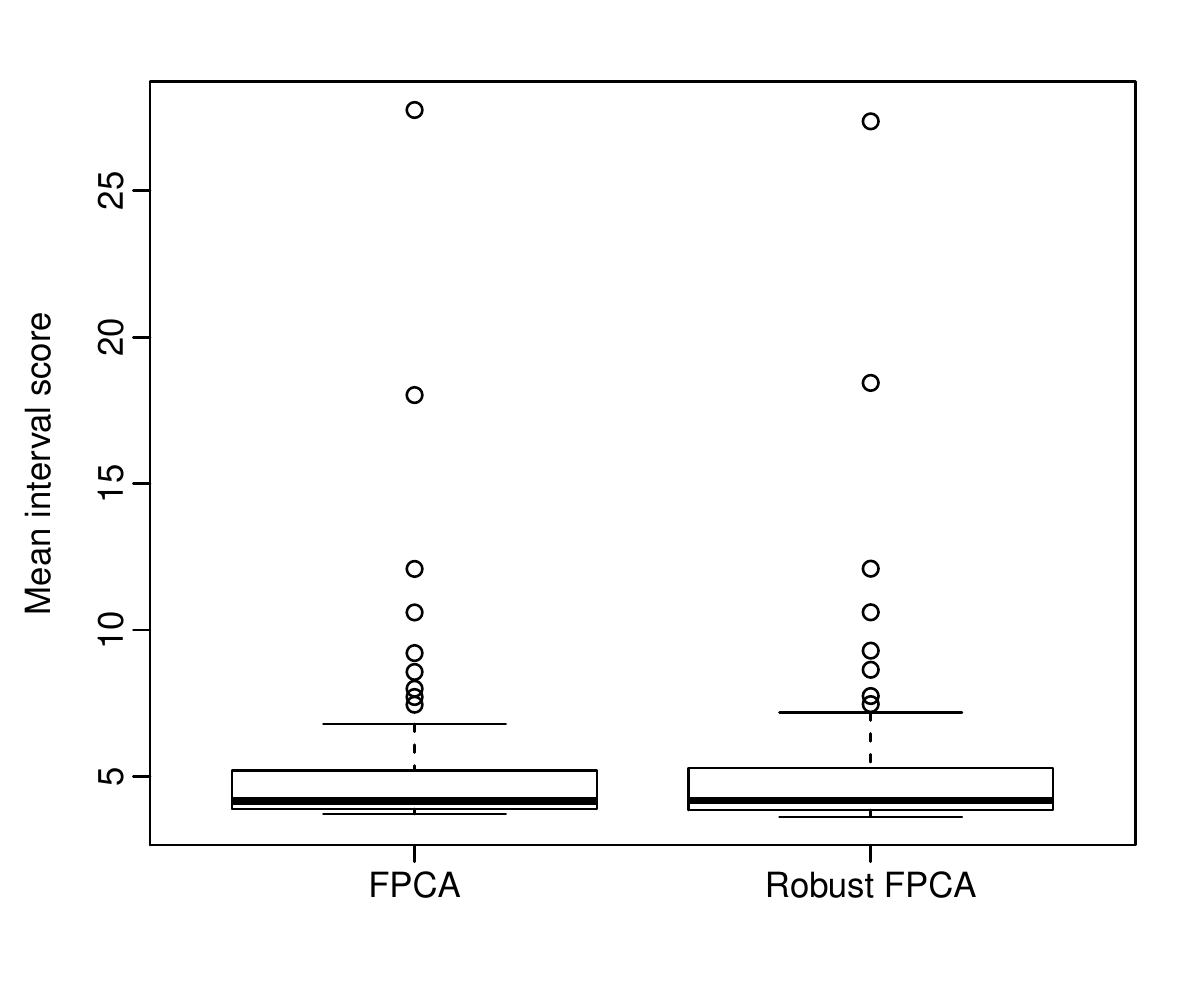}
\caption{Comparison between the standard and robust FPCAs for the point and interval forecast accuracies used in the TS method.}\label{fig:point_interval}
\end{figure}

Using the ARIMA forecasting method, we compare the point and interval forecast accuracies between the standard and robust functional principal component analyses (FPCAs) for the TS method. As shown in Figure~\ref{fig:point_interval}, we found that the robust FPCA provides slightly more accurate point and interval forecasts than the standard FPCA. 

Averaged over the last 72 days of the forecasting period, the averaged MSFE is 2.24, the averaged MAFE is 1.14, and the averaged mean interval score is 5.42, for the standard FPCA. For the robust FPCA, the averaged MSFE is 2.14, the averaged MAFE is 1.10, and the averaged mean interval score is 5.39. This result further confirms the advantages of applying a robust method in the presence of outliers.

\subsubsection{Updating point forecasts}

\begin{figure}[!htbp]
\centering
{\includegraphics[width=8.5cm]{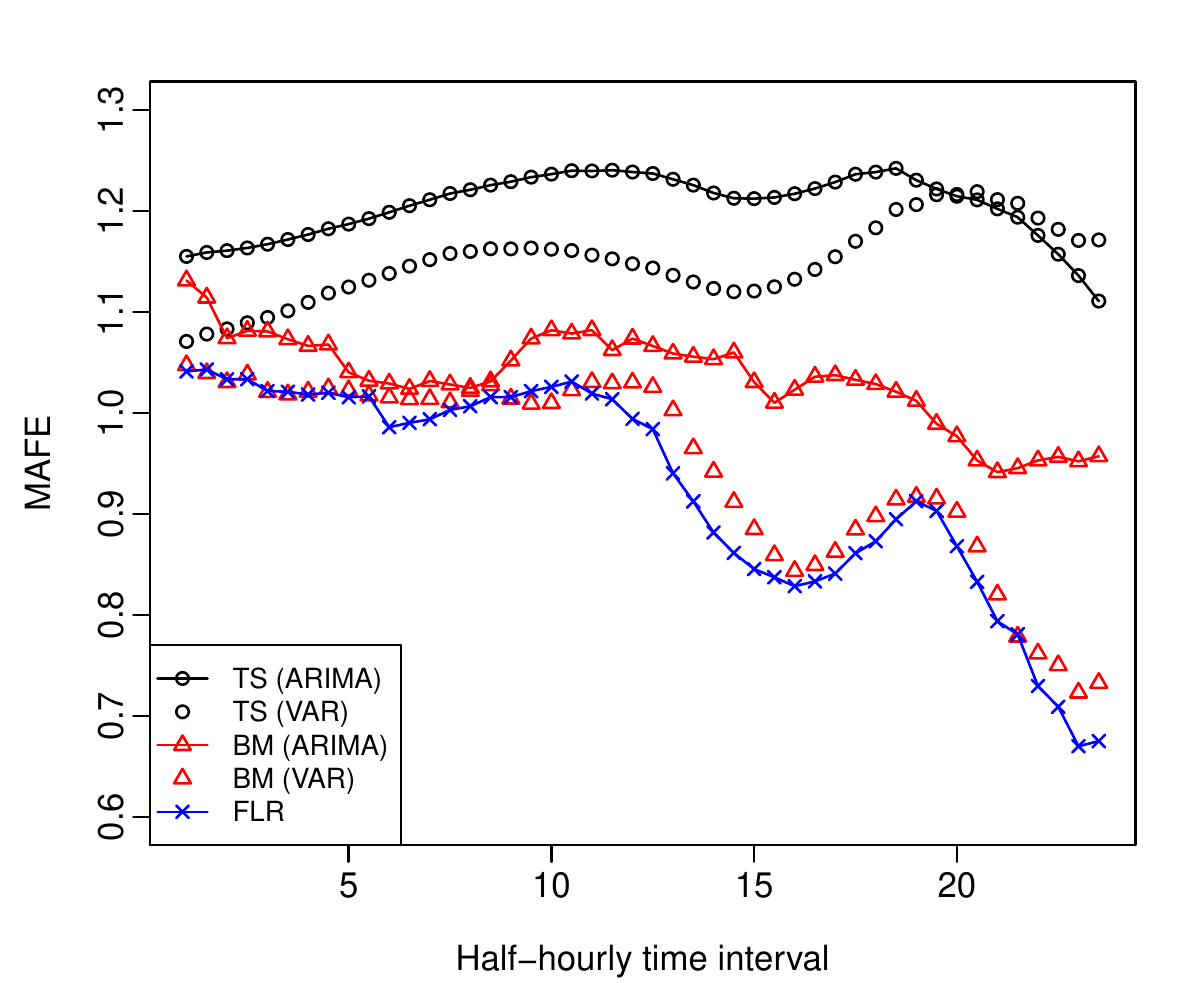}}
\qquad
{\includegraphics[width=8.5cm]{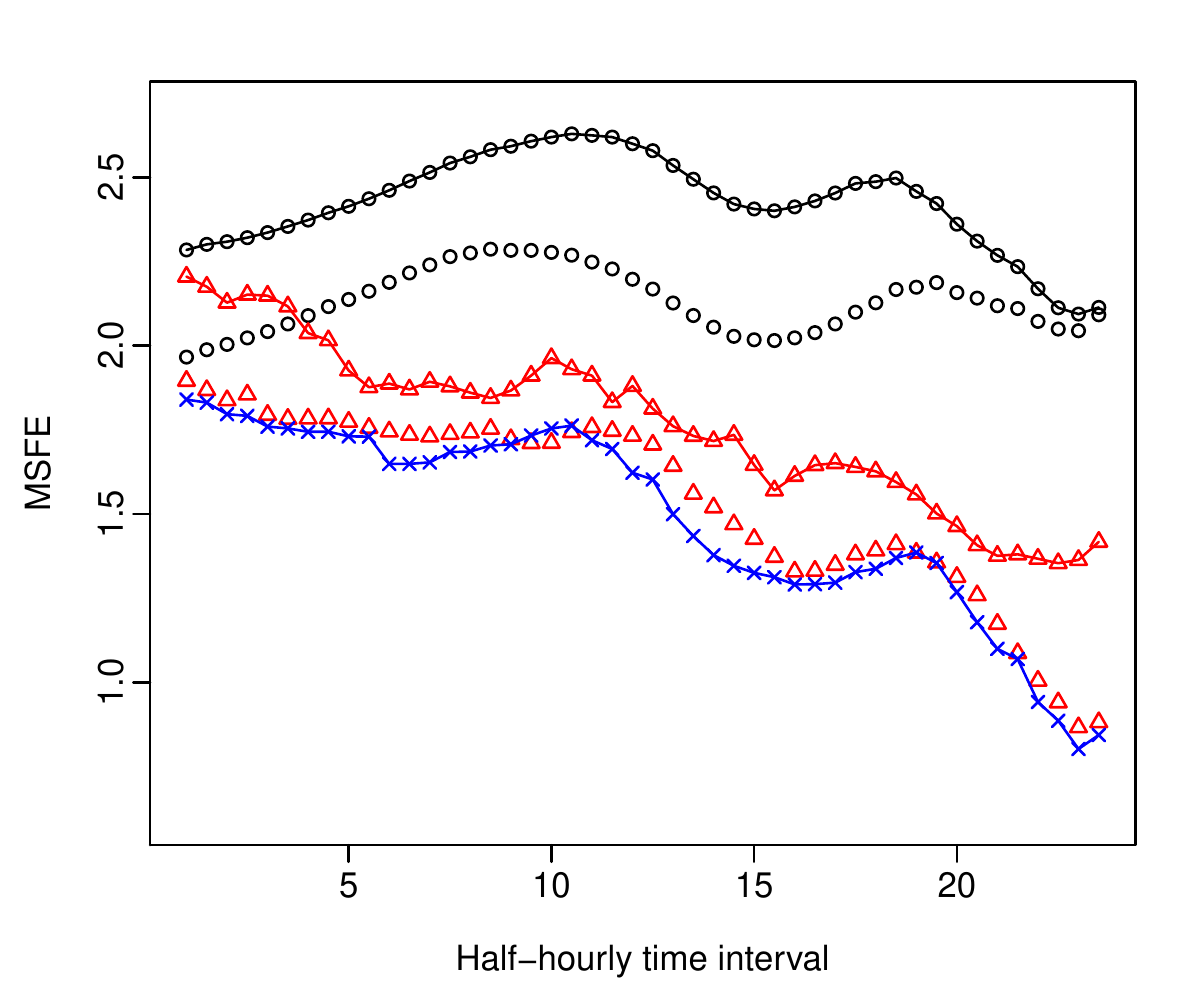}}
\caption{A comparison of point forecast accuracy, as measured by MAFE$_j$ and MSFE$_j$ for $j=3,\dots,48$, between the TS method and two dynamic updating methods.}\label{fig:dynamic_point_fore}
\end{figure}

When we have partially observed data in the most recent curve, we can dynamically update our forecasts in the hope of achieving better forecast accuracy. Two new dynamic updating methods have been proposed in Section~\ref{sec:4}, while a comparison of their point forecast accuracy is presented in Figure~\ref{fig:dynamic_point_fore}. The superiority of the two dynamic updating methods is evident from the reduction in forecast errors, when we observe more and more data points in the most recent curve. Averaged over the last 72 days of the forecasting period, we found that the FLR gives the most accurate point forecasts, as measured by both MAFE and MSFE over all discretized time points. To our surprise, the BM method with the multivariate time series forecasting technique performed marginally worse than the FLR method. The BM method gives the overall MSFE and MAFE as 1.54 and 0.95, respectively, whereas the FLR method gives the overall MSFE and MAFE as 1.49 and 0.93. Again, the VAR forecasting method produces more accurate forecasts than the ARIMA forecasting method for both the TS and the BM methods.

\subsubsection{Updating interval forecasts}

\begin{figure}[!htbp]
\centering
\includegraphics[width=8.8cm]{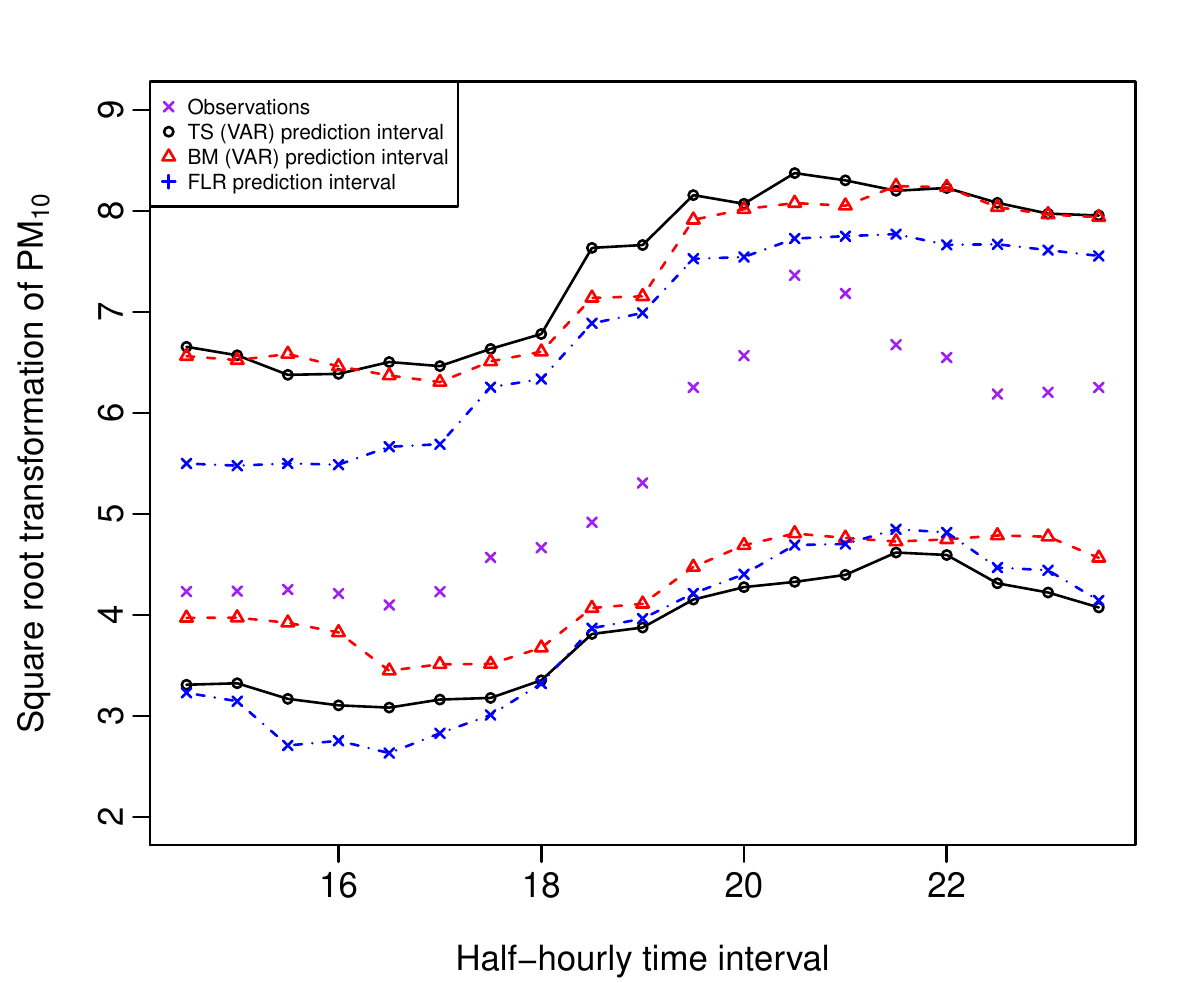}
\caption{With the partially observed data points from midnight to 2pm, we update the 80\% pointwise prediction intervals using the BM method and FLR between 2pm and midnight. For comparison, we include the 80\% pointwise prediction intervals obtained by the TS method.}\label{fig:PI_update_182}
\end{figure}

Suppose we observe the intraday PM$_{10}$ from midnight to 2pm, it is possible to dynamically update the pointwise prediction interval forecasts for the remaining time period of that day using the BM method and FLR. Since the BM method re-arranges the function support range, the updated interval forecasts can be obtained via the nonparametric construction of prediction intervals described in Section~\ref{sec:5.1}. Figure~\ref{fig:PI_update_182} presents the 80\% pointwise prediction intervals using these two methods, as well as the TS method. We found that the FLR produces the smallest mean interval score in all, and thus it provides most accurate evaluation of forecast uncertainty.

\begin{figure}[!htbp]
\centering
\subfloat[Standard FPCA]
{\includegraphics[width=8.52cm]{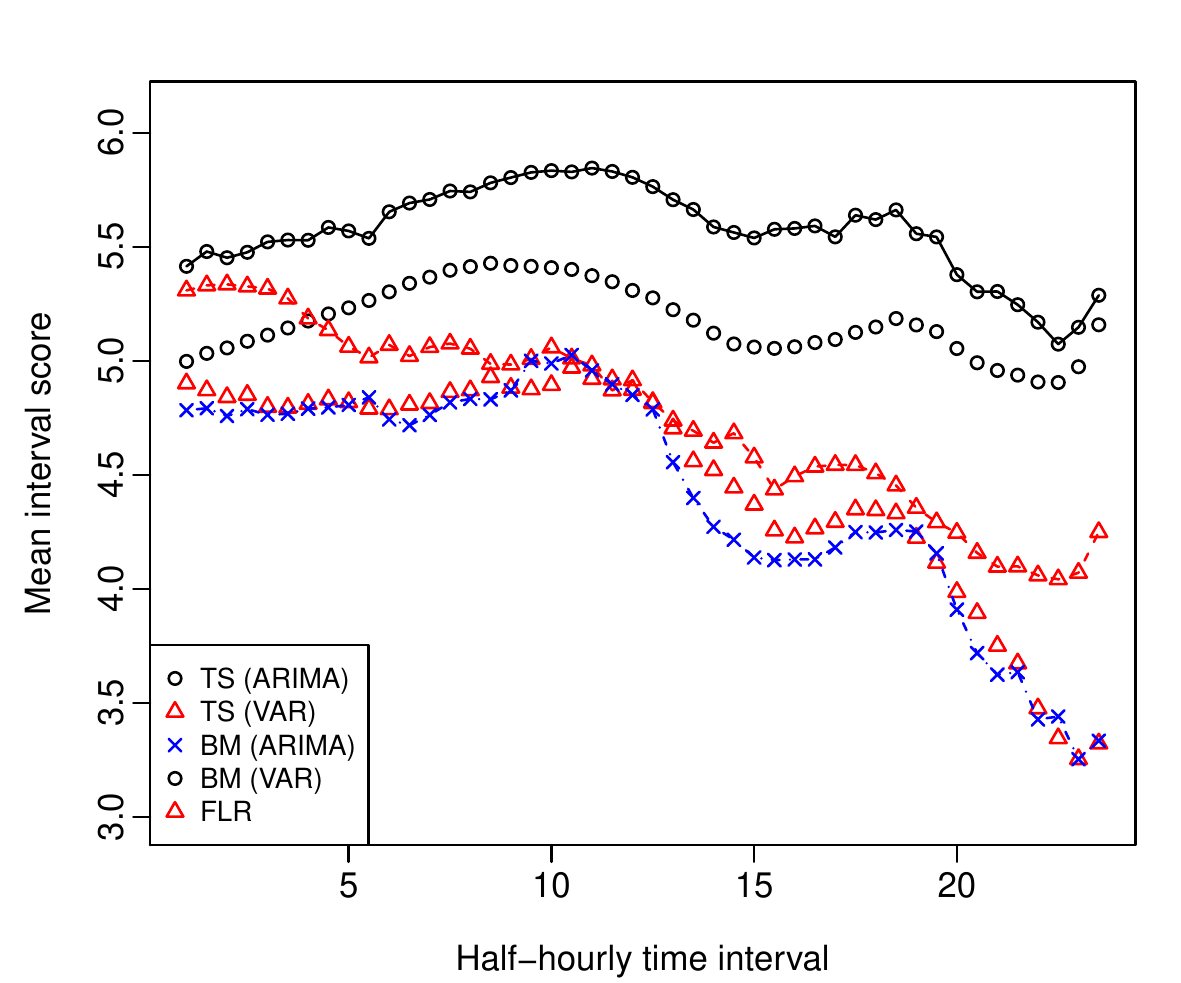}\label{fig:11a}}
\qquad
\subfloat[Robust FPCA]
{\includegraphics[width=8.52cm]{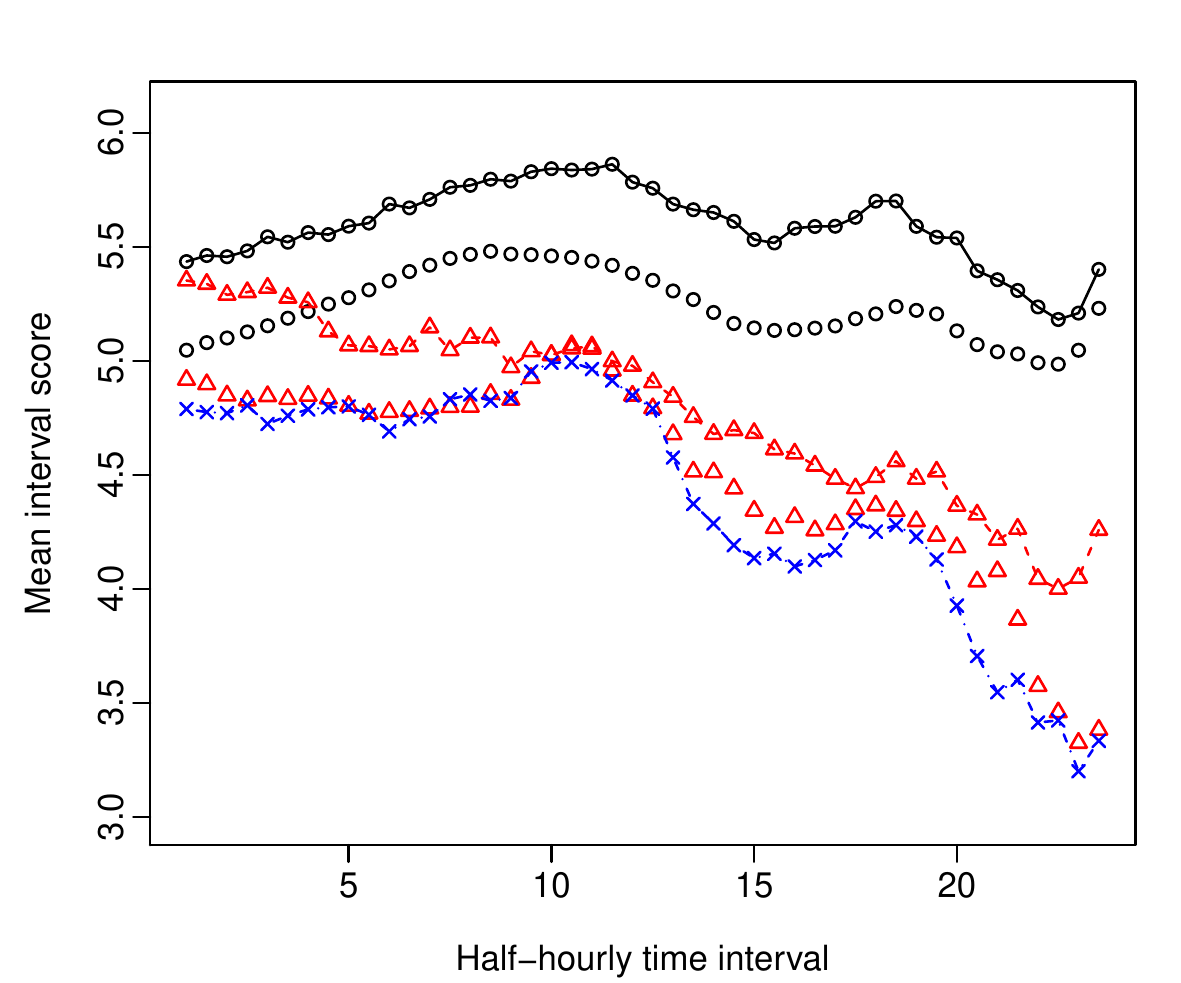}\label{fig:11b}}
\caption{Mean interval scores of the TS and BM methods with the ARIMA and VAR forecasting methods, and FLR at an 80\% nominal coverage probability.}\label{fig:11}
\end{figure}

Averaged over the last 72 days in the forecasting period, we show, in Figure~\ref{fig:11}, that the FLR produces the most accurate interval forecasts with the smallest averaged mean interval scores. The advantage of this method is highlighted by the fact that we sequentially observe more and more data points in the most recent curve. Between the standard and robust FPCAs, there is a small difference in terms of interval forecast accuracy for all the methods considered.

We compare the point and interval forecast accuracies averaged over different days in the forecasting period in the standard and robust FPCAs. As shown in Table~\ref{tab:1}, we found that the robust FPCA produces slightly smaller forecast errors than those obtained by the standard FPCA.

\begin{table}[!htbp]
\centering
\tabcolsep 0.37in
\begin{tabular}{@{}llccc@{}}
\toprule
FPCA method & Method & MAFE & MSFE & Mean interval score \\\midrule
FPCA 		& TS(uni) & 1.21 & 2.43 & 5.59 \\
		 	& TS(var) & 1.18 & 2.19 & 5.24 \\
	  		& BM(uni) & 1.04 & 1.79 & 4.80 \\
	  		& BM(var) & 0.96 & 1.58 & 4.52 \\
	  		& FLR       & \textBF{0.93} & 1.50 & 4.43 \\\midrule
Robust FPCA & TS(uni) & \textBF{1.17} & \textBF{2.37} & \textBF{5.57} \\
	      & TS(var) & \textBF{1.15} & \textBF{2.13} & \textBF{5.18} \\
	      & BM(uni) & \textBF{1.03} & \textBF{1.77} & \textBF{4.75} \\
	      & BM(var) & \textBF{0.95} & \textBF{1.54} & \textBF{4.48} \\
	      & FLR       & \textBF{0.93} & \textBF{1.49} & \textBF{4.42} \\\bottomrule
\end{tabular}
\caption{Comparison of the averaged MAFE, MSFE and mean interval score from the standard FPCA and robust FPCA. The smaller forecast errors are highlighted in bold.}\label{tab:1}
\end{table}

\section{Conclusion}\label{sec:8}

Our forecasting and updating methods treat the observed data as realizations of a functional time series, where the temporal dependency between the functional curves can be modeled by FPCA. As a by-product of using FPCA, the dimensionality of data is effectively reduced and the main features in the functional time series are represented by a set of functional principal components, which explain at least 90\% of the total variation in the half-hourly intraday PM$_{10}$ curves considered. The problem of forecasting the one-day-ahead intraday PM$_{10}$ curve has been overcome by forecasting $K$ number of retained principal component scores through a univariate or multivariate time series forecasting technique. Conditional on the historical curves, the estimated mean function and estimated functional principal components, the forecasts are obtained by multiplying the forecast principal component scores by the estimated functional principal components, and then adding the estimated mean function. 

When partial data in the most recent curve are sequentially observed, two new dynamic updating methods can update forecasts in order to improve forecast accuracy. The BM method re-arranges the function support range to obtain a complete data block, on which the TS method can still be applied. As an alternative to time series techniques, the FLR method first decomposes two blocks of functional time series corresponding to the partially observed data and remaining data periods via Karhunen-Lo\`{e}ve expansion, models the linear relationship between the two sets of principal component scores via OLS, and estimates the regression coefficient function from which the updated forecasts can be obtained. Based on the averaged MAFE and MSFE over different discretized time points in the forecasting period, the FLR method clearly shows the best point forecast accuracy of all the methods investigated.

As a means of measuring forecast uncertainty, we considered a nonparametric bootstrap method to construct pointwise and uniform prediction intervals for the TS and BM methods. Pointwise prediction intervals can also be updated through FLR, where the bootstrapped regression coefficient function and bootstrapped error function can be obtained by the maximum entropy bootstrapping and nonparametric bootstrapping, respectively. Based on the averaged mean interval score over different discretized time points in the forecasting period, the FLR method shows the best interval forecast accuracy in all methods considered. 

There are many ways in which the present paper can be extended, and we briefly mention three at this point. A natural direction for future research is to extend dynamic updating from one to multivariate functional time series. By capturing the correlation among multivariate functional time series, updating them jointly can possibly improve forecast accuracy more than updating each functional time series individually. Another possibility is to extend the VAR model to the VECM model, where the latter one can deal with non-stationarity among principal component scores. Instead of implementing a static FPCA method, a dynamic FPCA method is more appropriate for analyzing a functional time series and should bring potentially sizable improvements to the forecast performance. We leave each of these potential extensions to future research.

\newpage
\bibliographystyle{agsm}
\bibliography{dynamic}

\end{document}